\documentclass[journal]{IEEEtran}
\usepackage[utf8]{inputenc}
\usepackage[T1]{fontenc}
\usepackage{amsmath,amsfonts}
\usepackage{amsthm}%
\usepackage{mathrsfs}%
\usepackage{array}
\usepackage[caption=false,font=normalsize,labelfont=sf,textfont=sf]{subfig}
\usepackage{textcomp}
\usepackage{stfloats}
\usepackage{url}
\usepackage{verbatim}
\usepackage{graphicx}
\usepackage{cite}
\usepackage{listings}

\usepackage{xcolor}%
\usepackage{textcomp}%
\usepackage{manyfoot}%F
\usepackage{booktabs}%
\usepackage{algpseudocode}%
\usepackage{listings}%
\theoremstyle{plain}
\newtheorem{theorem}{Theorem}

\begin{document}

\title{Subspace Aggregation Query and Index Generation for Multidimensional Resource Space Model}

\author{Xiaoping Sun,
Hai Zhuge
  % <-this % stops a space
\thanks{Xiaoping Sun is with the Key Lab of Intelligent Information Processing, Institute of Computing Technology, Chinese Academy of Sciences, Beijing, China(e-mail: sunxiaoping@ict.ac.cn).}

\thanks{Hai Zhuge is with Great Bay University and Great Bay Institute for Advanced Study, Dongguan, Guangdong, China (e-mail: zhuge@gbu.edu.cn).}

\thanks{This work was supported by the National Science Foundation of China (project no. 61876048) and Multidisciplinary Innovation Research Group for New-Generation Intelligent Systems and Diagnostic-Therapeutic Applications (project no. 2025KCXTD031).}% <-this % stops a space
\thanks{Corresponding author: Hai Zhuge, email:zhuge@gbu.edu.cn}}

\markboth{Journal of \LaTeX\ Class Files,~Vol.~14, No.~8, August~2021}%
{Shell \MakeLowercase{\textit{et al.}}: An Article Using IEEEtran.cls for IEEE Journals}


\maketitle

\begin{abstract}
  Organizing large-scale resources in a multidimensional semantic space is an approach to efficiently managing and querying resources from different semantic dimensions. To support advanced applications, this paper proposes a resource space model for aggregation query on subspaces defined by a range within the partial order on the coordinate trees representing each dimension, where each point in the subspace contains resources aggregated along the paths of the partial order relations on the coordinate trees and the aggregated resources at each point can be measured, ranked and selected by applications. To efficiently locate non-empty points in a large subspace, an approach to generating graph index is proposed to build partial order relations on coordinates of dimensions to enable a subspace query to reach non-empty points through indexing links and aggregate resources along indexing paths to their super points.  Generating such an index is costly as the number of children of an indexing node can be large so that the total number of indexing nodes can be very large (exponentially growing with the number of dimensions and scale of dimensions).  The proposed approach adopts the following strategies to reduce the cost: (1) creating indexing nodes for non-empty points to link two coordinates at different dimensions, which can better reduce query processing cost while controlling the number of indexing nodes of the graph index; (2) the indexing nodes are created according to the probabilistic distribution calculated for estimating the costs of querying a non-empty point; (3) coordinates at one dimension having more resources are split by the index nodes of another dimension to balance the number of resources hold by indexing nodes; and, (4) shortcut links are added to connect sibling indexing nodes on a tree index to support an efficient query on linear order coordinates. Analysis and experiments show the effectiveness of the generated index in supporting subspace aggregation query. 
\end{abstract}


\section{Introduction}
\subsection{Motivation}
\IEEEPARstart{O}{rganizing} resources in a multidimensional semantic space can support efficient operations on a large set of resources from different semantic dimensions \cite{RN1}. The Resource Space Model (RSM) is created for managing resources in a multidimensional space where each dimension is represented by a classification tree on resources and finely classifies another dimension so that the space can be classified into exponentially smaller space when the number of dimension increases \cite{RN2}.  A multidimensional Resource Space (RS) consists of multiple abstraction dimensions, and every point is specified by coordinates at different dimensions of the space. Related research includes automatically extracting dimensions from a large set of resources \cite{RN3, RN4}, and multi-dimensional methodology for analyzing and modelling reality \cite{RN5}.

A partial order relation can be defined on the coordinates of each dimension so that a tree of coordinates can be formed and each path on the tree represents partial relations. Queries with such a partial order relation on the hierarchical coordinates at different dimensions are required by many advanced applications. 

In a resource space, coordinates at a dimension are often in a tree structure to represent inclusion and subclass relation between coordinates. Queries with such a structure of coordinates at a dimension can aggregate resources at different levels of coordinates at the dimension, which is different from queries on numerical values in classical multidimensional databases.  Subspace query is a general query to locate resources within a given ranges of dimensions in classical multidimensional data \cite{RN6}. A non-empty point of a subspace is an intersection of the resources represented by the coordinates at the dimensions of the subspace. Calculating intersections between resource sets specified by coordinates can be costly as the number of points increase exponentially with the increase of the number of dimension and the depth of the structure of each dimension. 

To support efficient query, a multidimensional index needs to be generated to index non-empty points, thereby improving query performance. However, classical multidimensional indexes cannot be directly used to process hierarchical coordinates and classical aggregation query concerns aggregative operation on the values of continue variables \cite{RN6, RN7}. The partial order relation on dimensions makes it difficult to use a constant number of indexing nodes to partition a resource space into a constant number of subspaces. The number of indexing nodes can be huge and selecting a proper set of non-empty points for indexing a resource space is a challenging problem.

This paper proposes a subspace query model and a graph index generation approach based on the coordinate trees of dimensions to implement subspace queries to efficiently aggregate resources within subspace.

\subsection{Representation of the Resource Space Model}

A Resource Space (in short $RS$) represented as $RS(X_1, ..., X_n)$, has n abstraction dimension $X_1, ..., X_n$ \cite{RN2}, each of which consists of a set of coordinates represented as $Xi = {c_i1,  c_i2, ..., c_ik}$, classifying resources into points in the space, so that each point can locate a set of resources by its coordinates. Each point has a projection (a coordinate) at each dimension. A point is respresented by $p = <c_1, ..., c_i> $ where $c_i$ is a coordinate of the $ith$ dimension $X_i$, identifying a set of resources for $c_i \in X_i (i = 1,...n)$.  Let $R(p)$ be the set of resources in (represented by) $p$.   If $|R(p)|>0$, $p$ is called a non-empty point.

For example, a two-dimensional resource space $RS(topic,date)$ manages scientific papers by a topic dimension with topic keywords as coordinates and a date dimension with date strings as coordinates. A point $p=<database, 2025>$ represents papers published in year 2025 on topic database. A partial order relation $\subset_{X_i}$ such as subclass relation and inclusion relation can be defined on a dimension to form a coordinate tree to represent inclusion, subclass and comparison semantics between coordinates within the dimension. For example, $\subset_{topic}$ defines the topic inclusion relation between coordinates of the topic dimension, the coordinate $database$ includes descendant coordinates such as $index$, $storage$ and $model$, and the coordinate index consists of $B-tree$, $R-tree$ and so on. <date defines that coordinate $2020$ at the date dimension includes month $2020-1$, $2020-2$ and so on.  Resources can be queried and organized according to partial orders on the coordinates.

A partial order relation between two points can be defined if partial orders occur between coordinates of the points. For example, point $<database, 2020>$ includes descendant points $<database, 2020-1>$, $<database, 2020-2>$, ..., etc. It also includes descendant points such as $<index, 2020-1>$, $<index, 2020-2>$, ..., etc.  The point $<database, *>$ includes descendant points such as $<database, 2020>$, $<database, 2021>$, ..., etc. 

A range $S_i=[l_i, u_i]\subset_{X_i}$ is defined on two coordinates $l_i$ and $u_i$ by a partial order $l_{i}\subset_{X_i}u_i$ at dimension $X_i$. $S_i$ contains the coordinates within the range of dimension $X_i$. A point $p=<c_{1}, ..., c_{n}>$ belongs to $S_i$ if $l_i \subset_{X_i} ci \subset_{X_i} u_{i}$.

\subsection{Examples of subspace query in Resource Space Model}

A subspace $RS(S_1 ..., S_n)$ is a set of points whose coordinates are within ranges $S_1$ ..., and $S_n$ of dimension $X_1$, ..., and $X_k$ in a resource space $RS(X_1, ..., X_n)$. A subspace query $sub(S_1 ..., S_n)$ obtains a set of non-empty points and their corresponding resources within $RS(S_1 ..., S_n)$. Points without resources should be excluded from the query result. For example, in resource space $RS(topic, date)$, a subspace query $sub(database, [2020-2021])$ obtains papers on topic dimension database and date dimension from $2020$ to $2021$. Papers at points including $<database, 2020>$ and $<database, 2021>$ are in the query result set. Subspace query $sub(database, *)$ obtains points $<database, 2020>$, $<database, 2021>$, ..., etc. The result of a subspace query $sub([index, database], 2020)$ contains resources at points with date coordinate $2020$ and topic coordinate defined by a range from $index$ to $database$, determined by the topic inclusion relations.

An aggregation operator agg can be applied to a range $S_i$ at dimension $X_i$ of a subspace query as $sub(S_1, S_2, ..., agg(S_i), S_n)$ to let each point contain its own resources and the resources aggregated from its descendant points according to partial order relations on dimension $X_i$. Partial orders induced by dimensions without aggregation operator should be excluded. The result is also a set of points within subspace $RS(S_1 ..., S_n)$ where each point is a collection of resources of all its descendant points. The following are three examples of such aggregate queries on resources space $RS(topic, date)$:

(1)	$sub(agg([index, database]_{\subset_{topic}} ), [2020, 2021]_{\subset_{year}})$. It specifies an aggregation operator within range $S_{topic}=[index, database] \subset_{topic}$ at $topic$ dimension by subclass relation $\subset_{topic}$. The range is restricted by two coordinates: $index$ and $database$. $S_{date}=[2020, 2021]_{\subset_{year}}$ is a range defined by two coordinates $2020$ and $2021$ on the temporal order relation $\subset_{year}$ at date dimension. $RS(Stopic, Syear)$ contains four points: $p1=<index, 2020>$, $p2=<index, 2021>$, $p3 = <database, 2020>$ and $p4 = <database, 2021>$. As the agg operator is specified only on Stopic, $p3$ contains papers at point $p3$ and point $p1$; $p4$ contains papers at point $p4$ and point $p2$. However, $<database, 2021>$ does not contains papers from point $<index, 2020>$ because the aggregator operator is not specified on Sdate although $index$ $\subset_{topic}$ $database$ and $2020 \subset_{year} 2021$. Resources are organized at each point within the subspace for further processing. Each point can have its own ranking and summarization results. For example, papers aggregated at point $<database, 2021>$ can be ranked and summarized according to their relevance and importance within database topic in year $2021$.  Similarly, papers at point $<index, 2021>$ can be used to select and summarize papers within the topic $index$ in year $2021$.

(2) $sub(agg([index, database]_{\subset_{topic}}), agg([2020, 2021]_{\subset_{year}}))$. It aggregates papers along partial order relations on two dimensions. Papers at point $<index, 2020>$ are aggregated into point $<database, 2021>$. Point $<database, 2021>$ contains all papers published from $2020$ to $2021$ on $database$ and index $topic$. Therefore, ranking and summarization at point $<database, 2021>$ are based on papers on database $topic$ and $index$ topic from year $2020$ to year $2021$.

(3) $sub(agg(Stopic=[none, database]_{\subset_{topic}}), agg(Sdate=[2020, 2021]_{\subset_{date}}))$. It aggregates papers from both topic dimension and date dimension. The range $S_{topic}=[none, database]_{\subset_{topic}}$ indicates that all coordinates under database are aggregated at the  database coordinate. For example, point $<index, 2021>$ contains papers at points $<B+-tree, 2020>$, $<KD-tree, 2020>$, $<B+-tree, 2021>$, $<KD-tree, 2021>$ and so on. Point $<database, 2021>$ contains papers from points such as $<index, 2021>$, $<model, 2021>$, and $<storage, 2021>$, and it contains papers from points such as $<index, 2020>$, $<model, 2020>$ and $<storage, 2020>$.

\subsection{Necessity and difficulty of implementing efficient subspace query in resource space}

Different from classical multidimensional database models, the RSM is designed to process queries on resources such as documents and multimedia files in a space with partial order relations on discrete coordinates of each dimension. A subspace query with aggregation operator is to aggregate resources into each point of different levels of partial order relations so that resources can be processed at the points determined by partial order relations on the coordinates. Classical aggregation operators in multidimensional database implement numerical calculations on values of data through aggregation. The underlying value does not need to be directly included into the points at higher level. In the result of a subspace query, each point corresponds to a set of resources like papers collected from its descendant points by following the partial order relations on the topic dimension and the date dimension. The points are organized by partial order relations and can be further operated such as ranking and summarization by applications. Thus, it is necessary to aggregate resources at points.

A resources space can be implemented based on either a relational database or a key-value document database. For example, a relational database table $RS$ can be used to store papers of a resource space $RS(topic, database)$. One row of the table corresponds to one paper.  The $topic$ dimension and the $date$ dimension are treated as two string attributes of the table. The $topic$ attribute of a row of a paper stores the keyword strings of the coordinate and the ancestor coordinates of the paper. Then, a subspace query $sub(database, [2020-2021])$ can be implemented by a SQL query: SELECT * FROM RS WHERE topic = 'database' AND date BETWEEN '2020' AND '2021', which includes papers in both point $<database, 2020>$ and point $<database, 2021>$.

Key-value document databases such as MongoDB or CouchDB can be also used to implement a resource space by using coordinates of a point as the key to store and query resources at the point. A MongoDB query can implement the above subspace query:

{\raggedright
{\small db.papers.find$(\{ $  }
}

{\raggedright
\hspace{15pt}{\small topic: 'database',}
}

{\raggedright
\hspace{15pt}{\small publicationDate: \textit{paper\_citation\_count}}
}

{\raggedright
\hspace{15pt}\hspace{15pt}{\small \$gte: new Date('2020-01-01T00:00:00Z')}
}

{\raggedright
\hspace{15pt}\hspace{15pt}{\small \$lt: new Date('2022-01-01T00:00:00Z')}
}

{\raggedright
\hspace{15pt}{\small $)\} $ ;}
}

SQL database and key-value database are optimized for processing keyword-based queries based on B-tree and hash index, which can efficiently implement subspace queries at a point and a range within one dimension.  However, when processing a subspace query, an SQL query needs to be executed for each point in the subspace, even for empty points.

When the number of points within a subspace is large, it requires large number of queries to be processed by either a SQL database or a key-value database, which makes it difficult to efficiently implement a subspace query.

For example, to process the above third example query based on the relational database table, a SQL query SELECT * FROM RS WHERE topic LIKE '%database%' AND date BETWEEN '2020' AND '2020' can obtain resources at the point $<database, 2021> $together with resources at its descendant points such as $<database, 2020>$ and $<index, 2021>$ but it does not tell whether there are resources at point $<index, 2021>$. Therefore, another query statement is needed to retrieve resources at point $<index, 2021>$, either from the result of last query or directly from the table, even there is no resource at the point. If a subspace query starts from the lower points, empty points still need to be separately queried. Similar problems occur when using key-value database queries or a keyword inverted index to obtain resources at points in a subspace. Moreover, when aggregating resources, resources at a point can be queried multiple times. Thus, processing a subspace query faces the challenging of improving query performance by avoiding querying empty points and reducing many times of accesses to resources. 

Building an index in a multidimensional data space is a classical solution to reduce the cost of accessing points outside the query range. However, partial order relations on coordinates of a dimension make it difficult to apply classical multidimensional index techniques to process a subspace query in a resource space with partial order relation on dimensions. Classical indexes such as B-tree, R-tree and KD-tree depends on a full order of coordinates of a dimension to partition a large subspace into smaller subspaces with a constant number of indexing links built for each indexing node. A subspace in a resource space cannot be partitioned into a set of smaller subspaces with a constant number of indexing links to each smaller one due to partial order relations on coordinates. For example, a string order can be applied to the keywords of the topic attribute of the table. Then, a string index can be built to help efficiently locate the coordinate \textit{database} of the topic attribute. However, the string order cannot be used to determine the sibling coordinate of \textit{database} according to the partial relation at the topic dimension. A query to obtain all descendants of \textit{database} needs to determine all the child coordinates of \textit{database} and exclude those sibling coordinates of \textit{database} like \textit{Artificial Intelligence}. There should be an indexing link to separate \textit{database} from its sibling coordinates, which makes it impossible to use constant number of indexing links of an indexing node to efficiently support querying \textit{database} and its descendant coordinates. The partial order relations on coordinates should be kept on an index.

A natural solution is to directly use the partial order relation on each dimension to generate a coordinate hierarchical tree as the index for this dimension so that resources with coordinates at this dimension can be efficiently retrieved through the coordinate tree. Resources at a non-empty point in the space can be obtained by finding the resource sets corresponding to each coordinate on the coordinate trees of dimensions and then calculating the intersection between resource sets of each coordinate. A graph index can be generated on a set of non-empty points where non-empty points are indexing nodes and partial order relations between points are indexing links between nodes. It enables the resources within a subspace to be retrieved by traversing the indexing links on the graph index without calculating intersections. Aggregation can be carried out along the index link, proceeding from descendant points to ancestor points. 

However, it is very expensive to index all non-empty points because the number of non-empty points in a resource space can be huge.  Thus, building such an index requires a strategy to select appropriate points to be indexed to support efficient query. 

\subsection{Contribution}

This paper proposes a subspace query for the multi-dimensional resource space created for organizing resources based on the partial order relations between points and generation of a graph-based index for supporting efficient query.  The major contributions include the following points: 

(1) Multi-dimensional aggregative query on partial order relations.  A subspace query with an aggregation operator is designed to query resources at points within a subspace of a resource space so that resources can be operated for various applications such as question answering and summarization at points organized according to the partial order relations in the resource space. It supports representation of different partial order relations in a subspace query with aggregation on resources in a multidimensional resource space.

(2) Optimized index generation with low cost to support efficient query.  A graph-based index is used to index non-empty points in a resource space for supporting efficient query. Each coordinate tree of a dimension is used to generate a tree index with coordinates as indexing nodes and partial orders on coordinates at the dimension as indexing links between nodes. The index can evolve with adding more indexing nodes between tree indexes of dimensions, and the partial order relation can be different types to represent the relations between various classes of resources, which can optimize the index structure to implement efficient queries at a relatively low lost for generating index.  A Mahalanobis distance is designed to estimate the cost of querying a non-empty point in the space. A probability distribution is designed on the Mahalanobis distance to add an indexing node with a probability based on the estimated cost of querying non-empty points so that those points with higher query costs can be indexed with higher probability. It can help control the number of indexing nodes while improving the query performance.  The indexing nodes are split with a probability to make the index balanced and the shortcut links are added between coordinates at the same level of partial order relations to help quickly locate the subspace.   

This study bridges the gap between classical index techniques and aggregate query of resources in a multidimensional resource space. Unlike previous index that depends on order relations on numeric values, the proposed graph index supports aggregative query on partial order relations on coordinates and points in the resource space. 

\section{Subspace Query with Aggregation in Resource Space}

Let $\subset_{X_i}$ be a partial relation on coordinates of dimension $X_i$, $*$ be the root coordinate of a dimension, and $p = <*, ..., c_i, *...>$ be a point in a resource space, simply represented as $p=<c_i>$ for ease of discussion, e.g., point $p=<database, *>$ in a two-dimensional resource space $RS(topic, date)$ is simply represented as $p=<database>$.  A partial order relation on coordinates of a dimension forms a coordinate tree where each node is a coordinate, and each link is a direct partial order between two coordinates. $c_i \subset_{X_i}^{k} s_i$ if $s_i$ has a path of length $k$ to $c_i$ on the tree of partial order relations on dimension $X_i$ and $c_i$ is a descendant coordinate of $s_i$. When $ci \subset_{Xi}^1 s_i$, $c_i$ is a child coordinate of $s_i$.

A partial order relation $p_1 \subset_{X_i} p_2$ between two points $p_1=<c_1, ..., c_n>$ and $p_2=<s_1, ..., s_n>$ in the space is induced if there is at least one coordinate $c_i s.t. c_i \subset_{X_i} s_i$ and $c_j = s_j$ or $c_j \subset_{X_i}^k s_j$ for the rest coordinates of $p_1$. $p_1 \subset_{X_i}^k p_2$ if $c_i \subset_{X_i}^k s_i$ or $c_i = s_i$ for all coordinates $c_i$, i.e., $p_1$ is a descendant of $p_2$. When $k = 1$, $p_1$ is a direct child of $p_2$. Specifically, $<c_1, ..., c_i, ..., c_n> \subset_{X_i}^k  <*, ..., ci, ..., *...>$, which can be simply represented as $<c_1, ..., c_i, ..., c_n> \subset_{X_i}^k <c_i>$. For example, $<index, 2020> \subset_{X_1}^1 <database>$, $<database, 2020> \subset_{X_2}^1 <2020>$, and $<index, 2020> \subset_{X_2}^1 <database, 2020>$. 

A range $S_i = [a_i, b_i]\subset_{X_i}$ can be defined at dimension $X_i$ according to the partial relation $\subset_{X_i}$ such that $a_i, b_i \in X_i$  and $a_i \subset_{X_i} b_i$. $p =<c1, ..., cn> \in S_i$ if $a_i \subset_{X_i} c_i \subset_{X_i} b_i$ for $i = 1,..., n$.  $none \subset_{X_i} c_i$ for all $c_i\in X_i$. So, $S_i = [none, b_i]$ contains all coordinates $c\subset_{X_i} b_i$ in $X_i$. $S_i = [none, *]$ contains all coordinates in $X_i$ and it can be omitted in a query statement. For example, $sub(S_{topic}=database, S_{date} = [none, *])$ can be represented as $sub(S_{topic}=database)$.

A subspace $RS(S_1, ...,S_n)$ = $\{p | p \in S_i \text{ for } i = 1, \ldots, n\}$ contains production of coordinates of $S_1$, ..., and $S_n$.  A subspace query $sub(S_1, ..., Sn)$ = $\{<p, R(p)>| p \in RS(S_1, ..., S_n) \text{ and } R(p) \neq  \emptyset\}$ locates non-empty points with corresponding resources at points in the subspace.  

An aggregation operator $agg(S_i)$ defined within a range $S_i = [a_i, b_i]_{\subset_{X_i}}$ at dimension $X_i$ is to aggregate resources at points within range $S_i$ and organize resources according to partial order relations between points:  $agg(S_i) =<V, E$>, where $V = {<p, A(p)>| \text{ for all } p = <c_1, ..., c_n>}$, satisfying: $p \in RS(Si)$ and $A(p) = \cup_s  R(s) \cup  R(p)$ for all $s =<s_1, ..., s_n> \in RS(S_i)$ s.t. $s_i \subset_{X_i} c_i$ and $s_k = c_k$ for $k \neq i$; $E = {<s, p> | s \subset_{X_i}^1 p\text{ and }s, p \in V}$.

A subspace query with aggregation defined on ranges $S_1$, $S_2$, $\ldots$, and $S_k$ at $k$ dimensions is defined as follows:

$sub(agg(S_1), ..., agg(S_k), S_{k+1}, ..., S_n) = <V, E>$, where $V = \{<p, A(p)>| \text{ for all } p = <c_1, ..., c_n>\}$, satisfying:  $p \in RS(S_1, ..., S_n)$, $A(p) = \cup_s  R(s) \cup  R(p)$ for all $s =<s_1, ..., s_n> \in RS(S_i)$  such that $s_i \subset_{X_i} c_i$ for $i =1, \ldots, k$ and $s_j = c_j$ for $j =k+1, dots, n$, $A(p) \neq \emptyset $;  $E = \{<s, p> | s \subset_{X_1}^1 p, s \subset_{X_2}^1 p, ..., s \subset_{X_k}^1 p \text{ and s }, p \in V\}$.

The query $sub(agg(S_1), ..., agg(S_k), S_{k+1}, ..., S_n)$ obtains resources at points within subspace $RS(S_1, ..., S_n)$ and aggregates resources from non-empty point $s$ to $p$ when $s \subset_{X_i} p$ at dimensions $_X1$, $X_2$, ..., and $X_k$ that have been used by aggregation operators in the query.  So not all dimensions need aggregation operations.

Let $q1 = sub(agg(S_1), ..., agg(S_k), S_{k+1}, ..., S_n)$ = $\{<t, A_1(t)>\}$ with $T$ = $\{t | t \in q_1\}$ and $q_2$ = $sub(agg(R_1), ..., agg(R_k), R_{k+1}, ..., R_n)$ = $\{<s, A_2(s)>\}$ with $S = \{s | s \in q_2 \}$. Then $q_1 \cap  q_2$ = $<V, E>$, where $V$ = $\{<p, A(p)>| p \cap S \text{ and } A(p) = \cup_s R(s) \cup R(p) \text{ for all } s =<s_1, ..., s_n> \in RS(S_i)\}$ such that $s_i \subset_{X_i} c_i$ for $i =1, \ldots, k$ and $s_j = c_j$ for $j =k+1, ..., n$, $A(p) \neq \emptyset $; $E$ =  $\{<s, p> | s \subset_{X_1}^1 p, s \subset_{X_2}^1 p, ..., s \subset_{X_k}^1 p \text{ and }s, p in V\}$.

 $q1 \cup  q2$ is a graph $<V, E>$, where $V = \{<p, A(p)>| p \in T\cup S \text{ and }A(p) = \cup_s R(s) \cup R(p) \text{ for all } s =<s_1, ..., s_n> \in RS(S_i) \}$ such that $s_i \subset_{X_i} ci$ for $i =1, \ldots, k$ and $s_j = c_j$ for $j =k+1, ..., n$, $A(p) \neq \emptyset$; $E =  \{<s, p> | s \subset_{X_1}^1 p, s \subset_{X_2}^1 p, ..., s \subset_{X_k}^1 p \text{ and } s, p in V\}$.

Different partial order relations produce different aggregation paths, e.g., $sub(agg(S_{date}=[*, 2021]_{\subset_{month-year}})$ aggregates resources from points $<2021-1>$, ..., $<2021-12>$ to point $<2021>$, and $sub(agg(S_{date}=[2020, 2021]_{\leq_{date}}))$ aggregates $<2020-1>$, $<2020-2>$, ..., $<2021-12>$ to point $<2021>$.

\section{GRAPH-BASED INDEX FOR SUBSPACE QUERY}

Generating a graph index for subspace query is to input the key-value database that stores the resources of a resource space and then output a graph index so that resources at any point $p$ can be efficiently accessed by $p$.

\subsection{Intersection Calculation for Processing Subspace Query}

A subspace query with aggregation is equivalent to calculating the intersections between the resource sets of points within each range at different dimensions specified by queries.

\begin{theorem}\label{thm1}
  Subspace query with aggregation operator follows the distributive law of the intersection operator: $sub(agg(S_1), agg(S_2), ..., agg(S_n))= \bigcap_{i=1...n}{sub(agg(S_i))} = <V, E>$ with $V =\cap_{i=1..n}{V_i}$ and $E = \cup_{i = 1,...,n}{E_i(V)}$, where $V_i$ is the points of $sub(agg(S_i))$ and $E_i(V)$ is the induced subgraph by $V$ in the query result of $sub(agg(S_i)).$
  \end{theorem}
\begin{proof}
    According to the definition of subspace query with aggregation, $sub(agg(S_i =[l_i, u_i]))$ contains points within $RS(S_i)$, and $sub(agg(S_1), agg(S_2), ..., agg(S_n))$ contains points in $RS(S_1, S_2, ..., S_n)$, which is an intersection of points in $RS(S_1), ..., RS(S_n)$. So, they have the same set of points and all partial orders on points within the results of the two queries should be included in the result of $q_1 \cap q_2$ (according to the definition of $q_1 \cap q_2$), which is equivalent to $sub(agg(S_1), agg(S_2))$. Nodes in $V$ are also included in $V_i$ of the result of $sub(agg(S_i))$ and the links induced by $V_i$ should be also used in query $sub(agg(S_1), agg(S_2), ..., agg(S_n))$ because it has aggregation operator within $S_i$. Thus, they have the same results.
\end{proof}

\begin{figure}[!t]
  \centering
  \includegraphics[width=0.8\columnwidth, keepaspectratio]{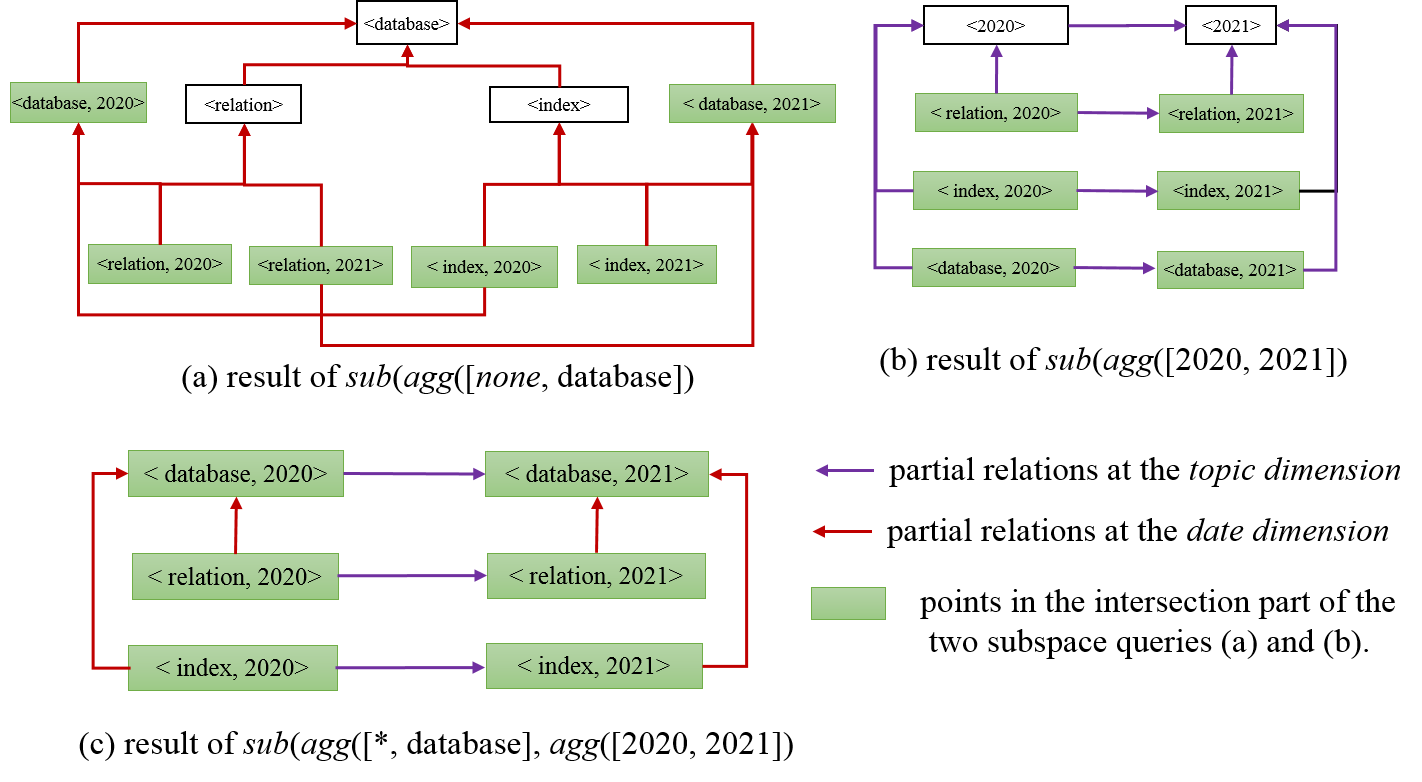}
  \caption{Subspace queries and results - graphs satisfying: (c) is the intersection of (a) and (b).}
  \label{fig_1}
\end{figure}

Fig.1 shows the results of the following three subspace queries, $sub(agg(S_{topic}=[none, database]_{\subset_{topic}}))$, $sub(agg(S_date=[2020, 2021]_{\subset_{date}}))$ and $sub(agg(S_{topic}=[none, database]_{\subset_{topic}})$, $agg(S_{date}=[2020, 2021]_{\subset_{date}}))$. The result of the first query contains a coordinate tree of the topic dimension. It has $RS([none, database], *)$ as its subspace and all descendant coordinates of database are included in the subspace, which contains points such as $<database>$, $<index>$, $<relation>$, ..., $<database, 2020>$, $<database, 2021>$, ..., $<relation, 2020>$, $<relation, 2021>$, ..., etc. The aggregation operation is conducted on the topic dimension. Then, the point $<database>$ includes resources from points $<index>$, $<relation>$, $<database, 2020>$, $<database, 2021>$, $<index, 2020>$, $<index, 2021>$ and so on. Although $<index, 2020>$ and $<index, 2021>$ are both in the target subspace, $<index, 2020>$ is not a child point of $<index, 2021>$ under the partial relation at the topic dimension. So, resources at point $<index, 2020>$ should not be included in point $<database>$ because the date dimension does not have aggregation operator.  Instead, resources in point $<index, 2020>$ should be contained in points $<database, 2020>$, $<index>$ and $<database>$. The result in the second query contains resources in points <2020> and <2021>. When an intersection is used by the two queries, the result contains the shared points of the two queries, including $<database, 2020>$, $<database, 2021>$, $<relation, 2020>$ and so on, which is equal to the result of the third query. Points such as $<database>$ and$ <2020> $are excluded because they do not belong to the subspace$ RS([none, database], [2020, 2021])$.

\subsection{Generating Graph Index for Improving Efficiency of Query}

According to Theorem 1, a subspace query on multiple dimensions can be decomposed into a set of subspace queries on one dimension. A tree index $G_i$ can be generated based on the partial order relation on coordinates at dimension $X_i$ for processing query $sub(agg(S_i =[l_i, u_i]))$.  Each dimension can have its own tree index and the non-empty points between points indexed by two tree indexes of two dimensions can be indexed by adding indexing nodes between two trees. Thus, the whole index is a graph consisting of nodes from tree indexes of dimensions and nodes between tree indexes.  

An index is generated after all resources are inserted into a key-value database using their coordinates as keys. Non-empty points can be enumerated through the key list of the key-value databases for adding indexing nodes and links, which forms a graph $G = <V, E>$, where $V$ is a set of indexing nodes and $E$ is a set of links between nodes. The approach to generating index nodes and links consists of the following two major steps, as described in Fig.2, where line 1-3 initializes the graph structure for storing the index.

(1)	Generate tree index $G_i$ for each dimension Xi by the following two major steps:

a)	For each non-empty point $p= <c_1, ..., c_n>$, create an indexing node $v$ with $p$ as its ID and link resources at point $p$ to $v$ and record the number of resources linked by $v$ (line 4-9). 

b)	For each coordinate $c_i$ at the dimension, create an indexing node $c_i$ and link it to $v$ by an inclusion link and insert the indexing node $c_i$ into the path from the root coordinate to $c_i$ on the graph index according to the coordinate tree (line 10-19). 

Within a tree index $G_i$, each indexing node is a coordinate $c_i$ corresponding to a non-empty point $<ci>$ and an index link $c_j \subset_{X_i} c_j$ represents a partial order relation between two non-empty nodes $c_i$ and $c_j$ and there is no coordinate cm s.t. $c_j \subset_{X+i} c_m \subset_{X_i} c_j \text{ with } |R(<c_m>)| >0$. That is, $G_i$ only contains non-empty points $<c_i>$ as indexing nodes. All non-empty points with coordinate $<c_1, ..., c_i, ..., c_n>$ are linked to the indexing node $c_i$ within $G_i$ for dimension $X_i$.  When a new resource is added to a point $p= <c1, ..., ci, ..., cn>$, if $c_i$ is not indexed by $G_i$, the indexing node $c_i$ is inserted and then the resource is linked to $c_i$.

(2)	Generate indexing nodes for points. After tree indexes are generated for all dimensions, indexing nodes on non-empty points are generated representing the intersections between existing indexing nodes on two tree indexes. An indexing node is linked to the indexing nodes on two trees $<c_i>$ and $<c_j>$ by two inclusion links to index the point $<c_i, c_j>$, which contains resources of the intersection between $<c_i>$ and $<c_j$>. Adding more indexing nodes between tree indexes helps reduce the cost of calculating intersections. However, it is impossible to index all such non-empty points, and it is a hard problem to generate an index with the optimal number of indexing nodes and low query costs. Thus, a probabilistic distribution is designed to select two nodes from two tree indexes of two dimensions to build new indexing nodes for non-empty points by calling procedure BuildIntersectionIndex in line 26. A procedure SplitNode is called in line 28 for making the index more balanced by splitting the nodes with great number of resources. A procedure AddShortCut is called in line 29 to add shortcut links within each indexing tree of dimensions to make a query on one dimension more efficient.  The procedure calls in Fig. 2 will be discussed in the following sections.

\begin{figure}[!t]
  \centering
  \includegraphics[width=0.8\columnwidth, keepaspectratio]{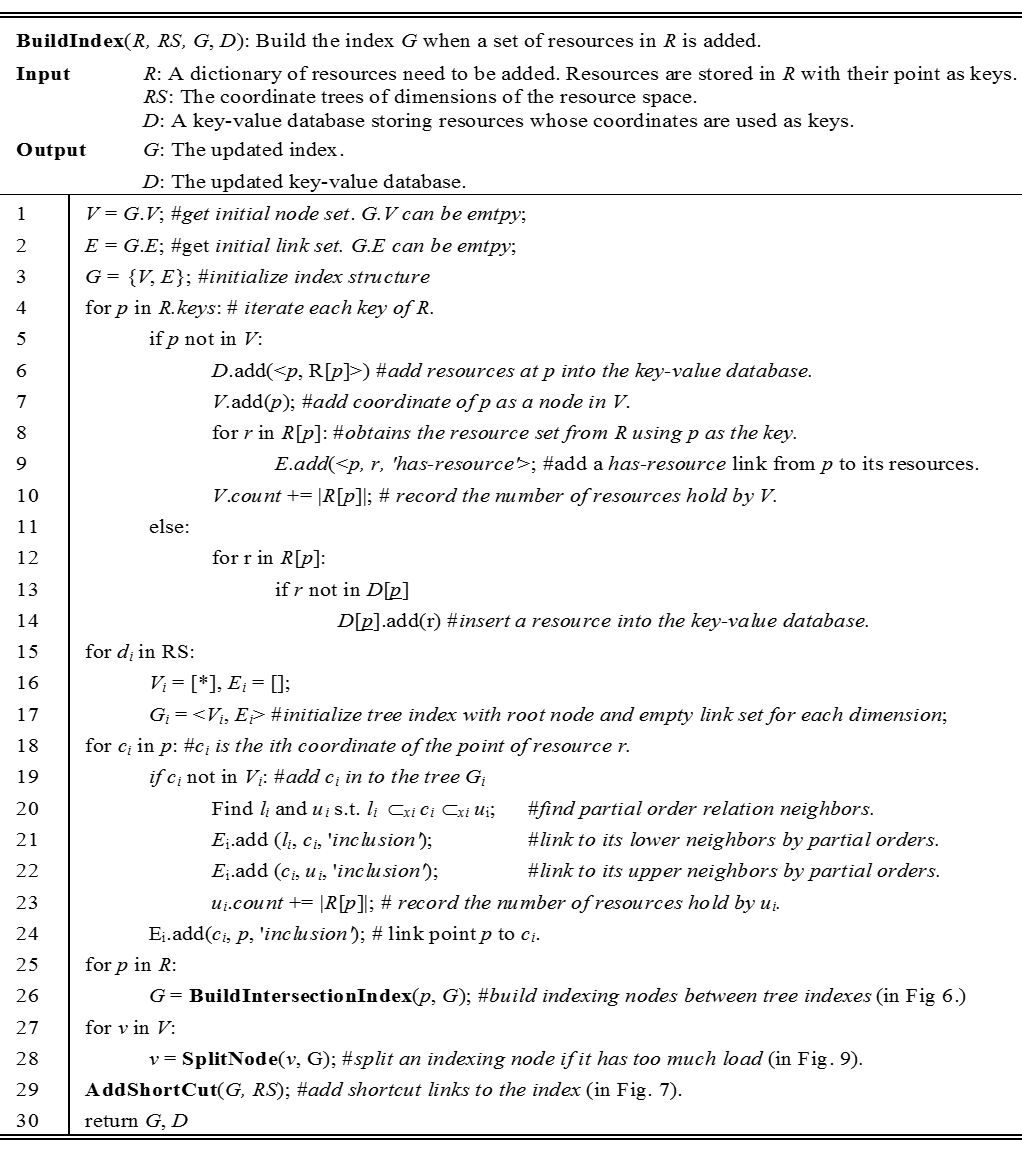}
  \caption{Main procedure for generating an index.}
  \label{fig_2}
  \end{figure}

A point query on a non-empty point $p = <c_1, ..., c_n>$ can be immediately processed by locating the indexing node $<c_1, ..., c_n>$ on the key value database using $<c_1, ..., c_n>$ as the key.  A subspace query is processed from the root coordinate of a tree index to locate non-empty points in the subspace. Fig. 3 contains a graph index consisting of two tree indexes, the left side is the topic dimension, and the right side is the date dimension. In the figure, each node uses its coordinate path in the dimension as the node ID. For example, $topic/CS/database/index$ is the node ID of the coordinate index in the topic dimension. A query on a point $<topic/CS/ database/index, date/2021/01>$ starts from the root of topic dimension to reach the indexing node $<topic/CS/ database/index, date/2021/01>$ shown as the block in green color at the lower left corner of Fig. 3. A query on $<topic/CS/ database, date/2021>$ can start from either the path on the two dimensions to reach the indexing node shown as the block in yellow color. Without the indexing node $<topic/CS/database, date/2021>$, it needs to calculate the intersection between two sets of resources obtained at indexing node $<topic/CS/database>$ and the indexing node <date/2021/01>.  Fig.3 shows an example of adding an indexing node on point $<topic/CS/database, date/2021>$ so that they can be accessed by following a path from the root of either tree index, which can reduce the costs of calculating intersection.

\begin{figure}[!t]
    \centering
    \includegraphics[width=0.8\columnwidth, keepaspectratio]{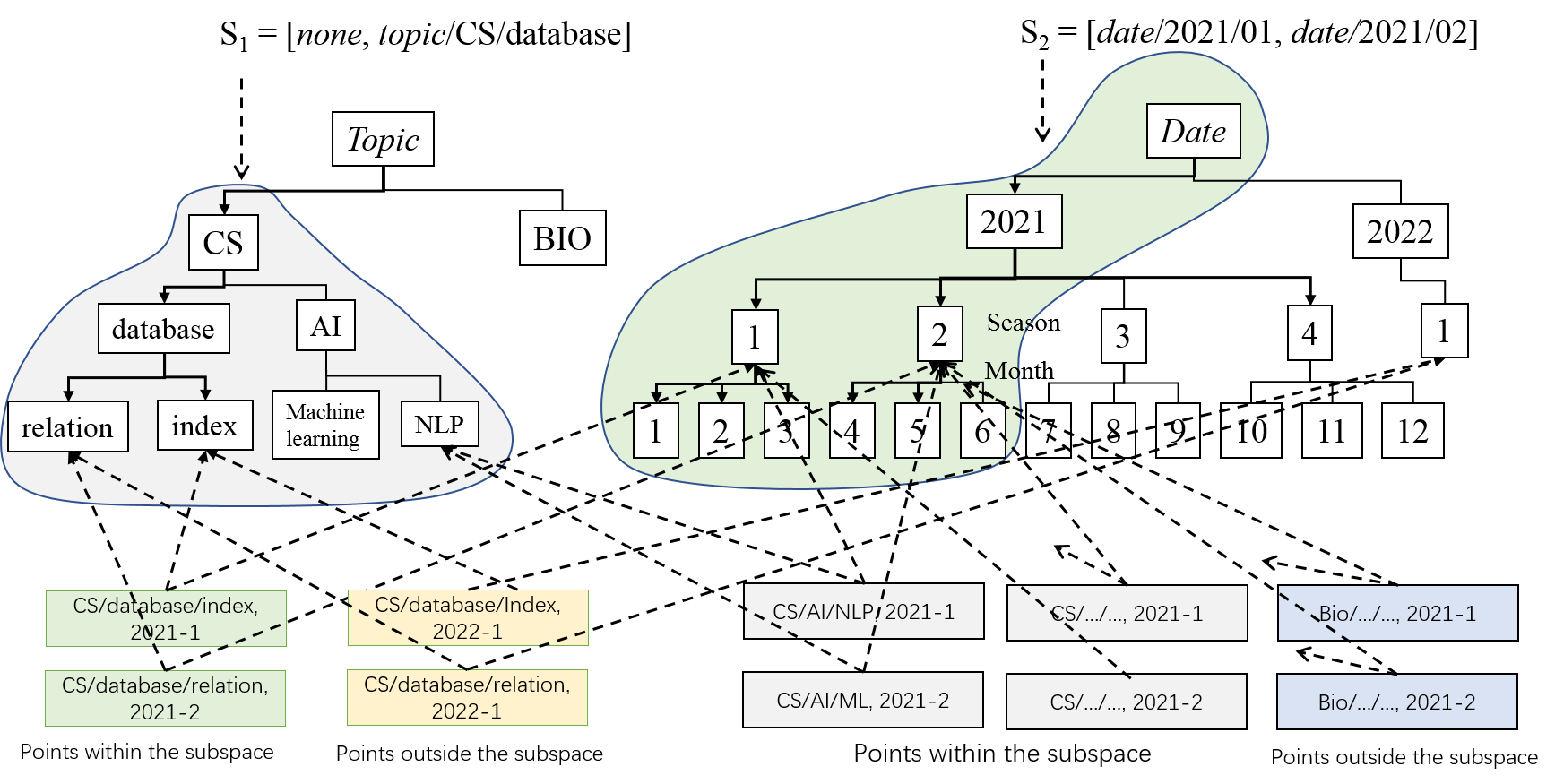}
    \caption{An index built on two dimensions.}
    \label{fig_3}
    \end{figure}
      
\subsection{Subspace query on graph Index}

A subspace query $sub(agg(S_1), ..., agg(S_k), S_{k+1}, ..., S_n)$ can be processed by separately processing $sub(agg(S_i))$, $sub(agg(S_j))$, ..., $sub(S_n)$ on index $G_1$, $G_2$, ..., and $G_n$. Each subspace query $sub(agg(S_i))$ can follow indexing links of $G_i$ from the root coordinate of the dimension to reach coordinates within the range specified on $X_i$.  The final query result can be obtained by calculating the intersection on the results of sub queries. For example, the indexing node $<database>$ holds all non-empty points with coordinate $database$ of the $topic$ dimension and the indexing node $<2020>$ holds all non-empty points with coordinate $2021$ of the $date$ dimension. Then, resources at point $<database, 2021>$ can be obtained by calculating the intersection between points indexed by $<database>$ and points indexed by $<2021>$. If the non-empty point $<database, 2021>$ is also indexed by an indexing node added by step (2), resources can be directly accessed without calculating intersections. 

A subspace query on a subspace $RS(S)$ with $S=\{S_1 = [l_1, u_1], ..., S_n=[l_n, u_n]\}$ on dimensions can be processed on the graph index in a greedy way:

\begin{list}{}{}
  \item{(1) Starting from the root indexing node of the tree at each dimension $X_i$, traverse the tree to locate each coordinate indexing node within range $S_i$.} 
  \item{ (2)	Check each coordinate indexing node $c_i$ within the range to locate links within the range.  }
  \item{(3)	For each link $l$ within $RS(S)$ from $c_i$, following the link $l$ to find the non-visited indexing nodes within $RS(S)$.}
  \item {(4) For each non-visited indexing node within the subspace, record the points within $RS(S)$ in the result, mark the indexing node as visited. }
  \item {(5) Follow partial relations like the inclusion links to reach child indexing nodes containing $RS(S)$ and locate points within $RS(S)$ for each child indexing node.}
  \item {(6)	For each point recorded in the result set, copy resources to their parent coordinates through the partial relations of each dimension if an aggregation operator is applied on the dimension. During the aggregation process, the links between points within the range are built to form the graph of the query result.}
\end{list}

\begin{figure}[!t]
  \centering
  \includegraphics[width=0.8\columnwidth, keepaspectratio]{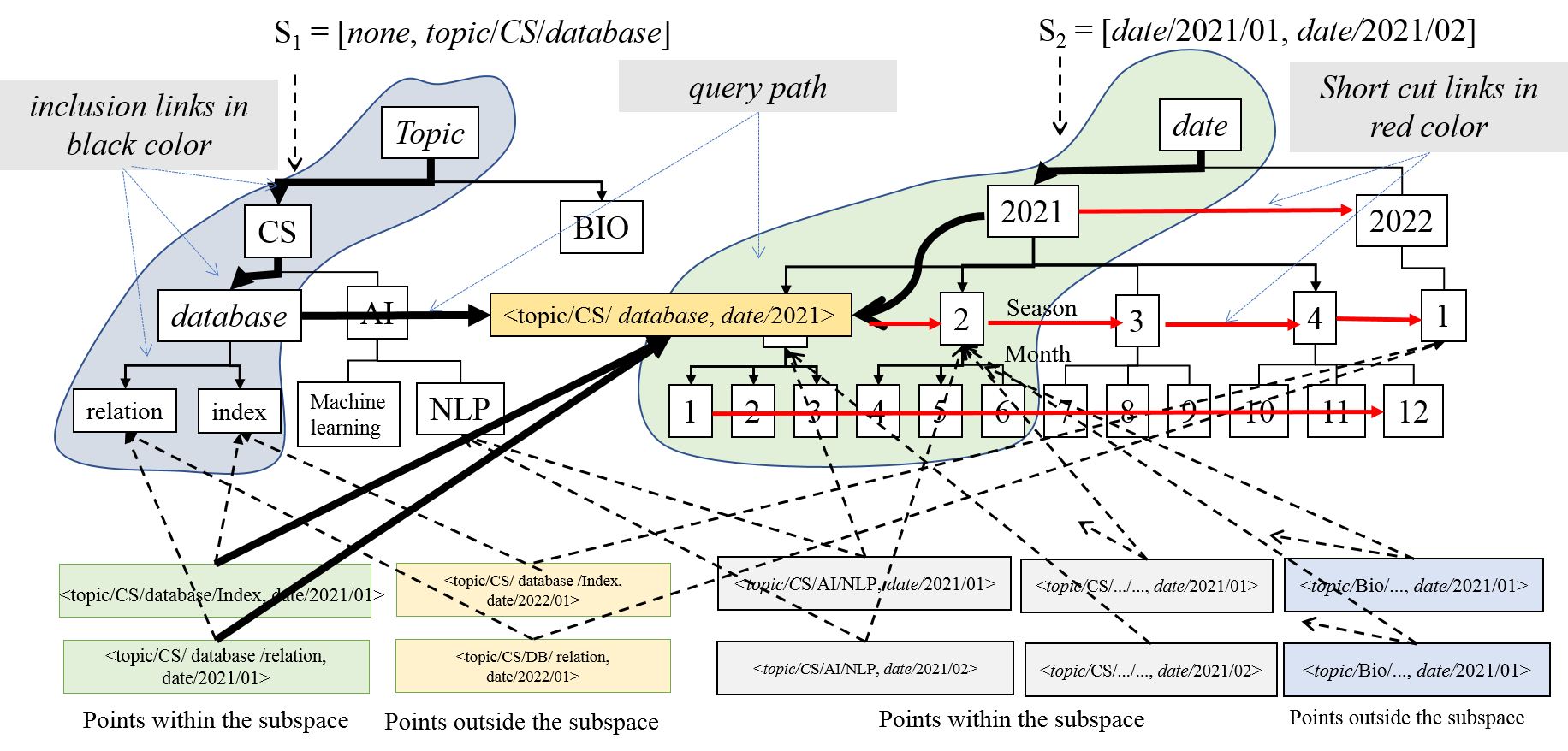}
  \caption{A query following links of the graph index.}
  \label{fig_4}
  \end{figure}
    
  Fig. 4 illustrates a subspace query on two ranges within the topic dimension and the date dimension. The coordinates within each range are marked by the irregular regions. $topic/CS/database$ and $date/2021$ represent the paths from the roots to the coordinates of the topic dimension and the date dimension respectively. The query paths are in bold black color. The query processing starts from the root indexing node at the two dimensions respectively. An indexing node $<topic/CS/database, date/2021>$ can be reached from either tree, and points within the range can be obtained through the links from the indexing node $<topic/CS/database, date/2021>$. Only two points within the range are visited from the links to the indexing node $<topic/CS/database, date/2021>$. If there is no such links, the query processing needs to conduct intersection calculations on resource sets at six points.

  \begin{figure}[!t]
    \centering
    \includegraphics[width=0.8\columnwidth, keepaspectratio]{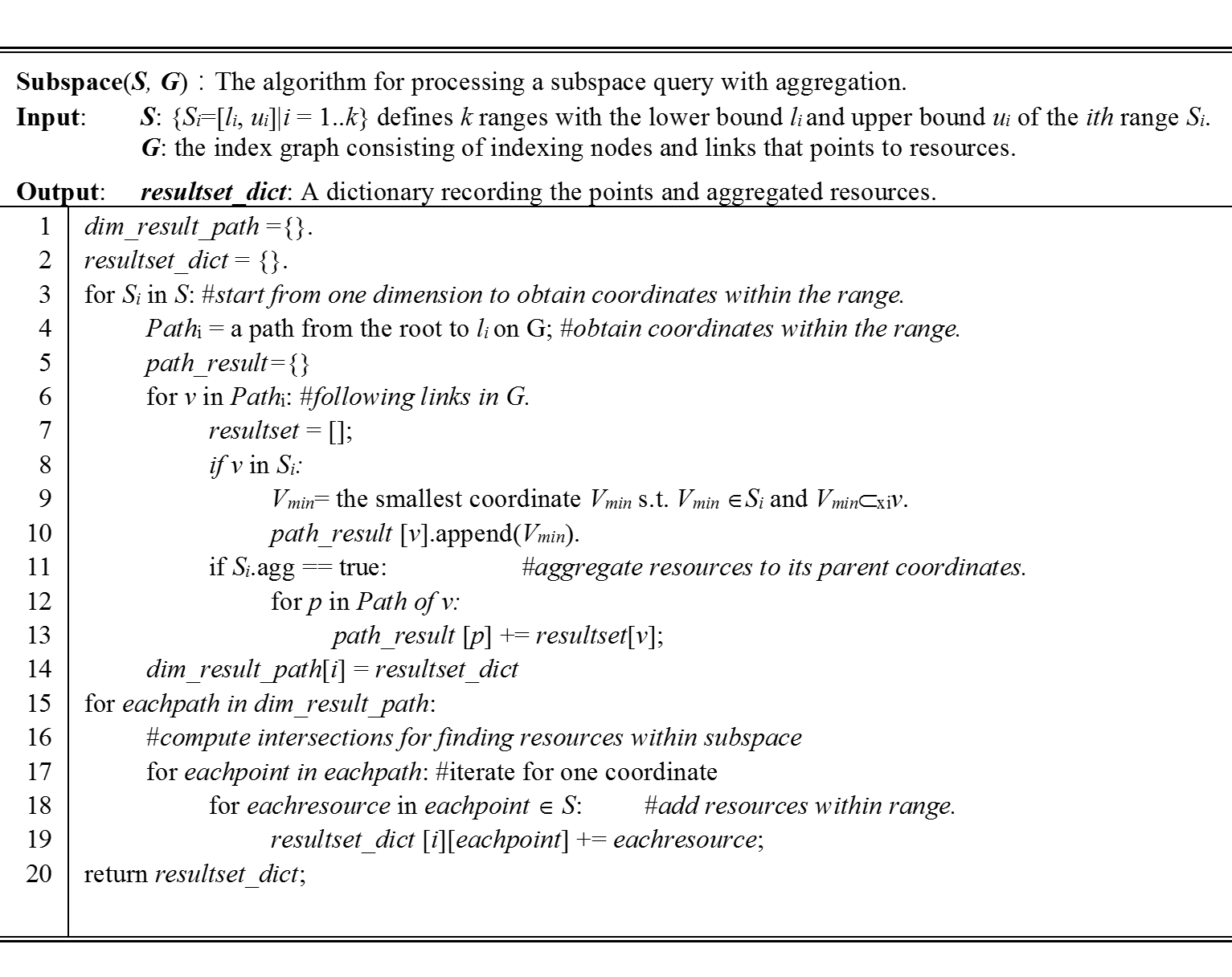}
    \caption{Subspace query supported by the graph index.}
    \label{fig_5}
    \end{figure}
    
Fig. 5 shows the algorithm for subspace aggregation query, where lines 1-15 scan each range $S_i$ specified in the query, lines 6-14 scan each coordinate on a path from the root to the lower bound of the range at a dimension to find the points within the querying subspace, lines 6-10 follow the indexing links of each coordinate on the path to find the smallest coordinate within $S_i$, lines 11-13 aggregate resources on the path, lines 15-19 calculate the intersection for locating resources within the subspace, and line 19 returns the query result.

\begin{theorem}\label{thm2}
  The traversing process starting from one dimension along inclusion links and ancestor links on $G$ can completely reach points within subspace $S$. 
  \end{theorem}
  \begin{proof}
    It can be derived from the generation steps. The index $G$ generated by the procedure BuildIndex presented in Fig. 2 can be regarded as a set of trees connected by non-empty points with resources. Thus, a traversing process will only cover those non-empty points within the querying subspace $S$.  A query can traverse the coordinate index on one dimension to obtain all the points within the range of this dimension. Then, each of these points is filtered using the ranges on other dimensions in the query so as to acquire all the points that meet the query range conditions. Following the links to points, all coordinates within all dimensions can be accessed for aggregation.
  \end{proof}

  A query on a graph index can be regarded as a search process that tests each coordinate of each resource one by one to check if it is within a range Si. The search process is speeded up through the index that skips empty points and reduce searching subspace.

\section{DIFFICULTY OF BUILDING AN EFFICIENT INDEX}  

Creating more indexing nodes on non-empty points to connect coordinate trees on different dimensions can make query more efficient. However, an index should have a bounded number of indexing nodes to achieve efficient query process, otherwise the cost for building index can be very expensive. This feature can be achieved by a classical index where each indexing node has a constant number of child nodes and each of its child nodes can help narrow down the search space until reaching the target. This cannot be achieved in a $RS$ with partial order relations because there is no partial order relation between child points of an indexing node. For example, $topic/CS/database$ can have ten child coordinates, each having a partial order relation with $topic/CS/ database$. This cannot be represented by two links because child points are independent. However, in a metric space, child points of an indexing node can be represented by two bounds. For example, an indexing node has many child points within an integer range $[0, 100]$, but it can just use two sub-links, one pointing to $[0, 50]$ and the other to $[50, 100]$ to partition the child points because half of child points of $[0, 100]$ can be represented by a range $[0, 50]$ due to the linear full order relation on coordinates. Thus, an index with constant indexing links can be built in a metric space. Although a full order can be built on coordinates, it cannot represent partial order relations. Thus, building an index in a resource space with partial relations on points needs to balance the index size with query processing efficiency. There can be exponentially non-empty points in a resource space, so it is impossible to index them all, therefore it is necessary to select a set of non-empty points to build an index to implement efficient query with bounded number of indexing nodes. 

The cost of a query on the index can be estimated by the number of resources visited on the path of the index during processing a subspace query because when a query is processed at an indexing node, it needs to locate the target point from all child points it has. If the indexing node holds exact target points within the querying range, the intersection calculation can be omitted for the points. If the indexing node holds points outside the target subspace, it needs to select its child points, which is equivalent to the intersection calculation on the points from two different dimensions. Thus, the cost of processing a query from a root point to a non-empty point on the index can be estimated by the summation of the numbers of child nodes on the path where each indexing node has a link to one of its child points that holds the target point until reaching the final target point. 

As there is an exponential number of non-empty points in a resource space, selecting a part of the points as indexing nodes for optimal query is a NP-hard problem.

\begin{theorem}\label{thm3}
  Generating an index with no more than n nodes on partial order relations at each dimension of a resource space to achieve the minimum querying cost for subspace query is an NP-hard problem. 
  \end{theorem}
  \begin{proof}
    In a Resource Space, each dimension $X_i$ has a tree $G_i$ of partial order relations between coordinates. Given a point
    $p=<c_{1}, ..., c_{n}>$ that has resources, $c_{i}$ has a path $P_{i}$ to its root of the dimension tree $G_{i}$. Then, the ancestor points of $p$ are the set $U=\{s\vert{}s \in {}2^{P1 \times P2 \cdot\cdot\cdot \times Pn}\}$. Obviously, we cannot build an index link from each \textit{s} of \textit{U} to \textit{p}. For a group \textit{D} of such non-empty points, each point \textit{p$_{i}$ }will have such a \textit{U$_{i}$} that contains all ancestors of \textit{p$_{i}$}. As each \textit{s} in \textit{U$_{i}$} also has inclusion relations with others in \textit{U$_{i}$}, a lattice graph on \textit{S$_{i}$} is formed and all \textit{S$_{i}$ }forms a
    larger lattice \textit{G$_{S}$ }from the root \textit{r} of the space to each \textit{p$_{i}$} in \textit{D}. When an index is built with inclusion relations, it should consist of all points directly holding resources and the root \textit{r} of space, together with a subset \textit{S}=\{\textit{s$_{i}$} $\vert{}$\textit{s$_{i}$} $\in{}$\textit{G$_{S}$} \} to form a sub-lattice graph \textit{T} with \textit{r} being the root, \textit{p}$_{i}$ being the leaf nodes, \textit{s$_{i}$ }being the internal indexing nodes and each link of \textit{T} represents an inclusion relation between the two end nodes. The cost of searching for a point \textit{p$_{i}$ }is the sum of the numbers of child links of points on a path from \textit{r} to \textit{p$_{i}$}. If each coordinate tree has a bounded depth, the searching depth on the index \textit{T} is also bounded by the depth of the deepest tree among all \textit{G$_{i}$}. Thus, the cost of building the index and processing a query is determined only by the number of indexing nodes $\vert{}$\textit{T}$\vert{}$ and the cost of determining the child node at each indexing node on the path leading to \textit{p$_{i}$}, i.e.,
     $cost(T) = \sum_{P_k\subset T}\sum_{s_i\in P_k}{|s_i|}$. Then, building such an index \textit{T} to obtain the optimal subspace query costs with bounded number of indexing nodes, $|T| \leq O(n^c)$ with $n = |D|$ can be modeled as $argmin_{\substack{T \subset G,\ 0\le|T|\le O(n^c)}}{\sum_{P_k\subset T}\sum_{s_i\in P_k}{|s_i|}}$, which is a hard problem. 
    
    This can be modeled as the constrained shortest path problem that tries to find a path from a source node \textit{s} to a target node \textit{t} with bounded resources and optimal cost on the weighted graph. Specifically, each parent node in $G_{s}$ has a cost that can be obtained by summing the number of nonempty points along the paths to the root of the space so that each link has a number of resources summed from its child point to its parent point in the space. Then, the cost $cost_{ij}$ of a link from \textit{i} to \textit{j} can be defined as $cost_{ij} = 1 - \vert{}R(l_{ij})\vert{}/\vert{}R(j)\vert{}$, where $R(l_{ij})$ represents the resources accumulated from point \textit{i} to \textit{j}. If $\vert{}R(l_{ij})\vert{} = \vert{}R(j)\vert{}$, $c_{ij} = 0$. 
    
    As \textit{T} contains a path from \textit{r} to each non-empty point $p_{i}$, the path $P_i$ from $r$ to $p_{i}$ with the minimal cost $\Sigma{}_{c_{ij}\in{} P_i}{c_{ij}}$ will also make the cost(\textit{T}) minimal. That is, if we can find a weighted shortest path from \textit{r} to $p_i$, we can make cost(\textit{T}) minimal by combining all such paths into \textit{T}, where each path from \textit{r} to $p_i$ has the minimal cost. Moreover, the weight $w_{ij}$ of each link $l_{ij}$ is equal to 1. That is, all links has the same weight. Then, $\vert{}T\vert{} =\Omega{}(\Sigma{}_{w_{ij}\in{}T}w_{ij}) \leq{} n^c $ indicates that the total weight should be bounded. That is, if we can build an index \textit{T} with bounded size and optimal costs within polynomial time, we can obtain a constrained weighted shortest path for each pair of \textit{r} and \textit{p$_{i}$}, which is an NP hard problem even when \textit{G} is a DAG \cite{RN8}.
  \end{proof}

\section{ADDING INDEXING NODES TO CONNECT COORDINATES}
\subsection{Resource Space}

The following two heuristic rules are for creating an indexing node to connect coordinates at two dimensions to help improve query efficiency with bounded number of indexing nodes. 

\textbf{Heuristic Rule 1}. The greater the difference in the number of resources between two coordinates, the higher the probability of adding an index node to connect the two coordinates. 
This rule can reduce the calculation of intersection between coordinates with a relatively small number of resources otherwise each resource of a coordinate with fewer resources needs to be compared with each resource of a coordinate with more resources, which significantly increases the cost of calculating intersection, especially for coordinates with fewer resources.  
\textbf{Heuristic Rule 2}. If the two coordinates are at different levels of tree indexes at two dimensions, an indexing node is built to connect them with a higher probability.
 This is because the two coordinates have a higher probability to hold a highly different number of resources than those at the same level.  

 When adding an indexing node, a Mahalanobis distance between two coordinates is used to implement the two heuristic rules with a probabilistic distribution for adding indexing nodes. The detailed procedure is as follows:

(1) For each coordinate $c_i$ on the path from the root at one dimension to $v$, locate one coordinate $c_j$ on the path at another dimension. 

(2) Build weight vectors representing the level and the number of resources of $c_i$ and $c_j$ respectively.

(3) Calculate a Mahalanobis distance between the two weight vectors as an indicator to measure the cost of calculating the intersection between the two coordinates.  

(4) Calculate a probability of creating an indexing node to connect two coordinates of different dimensions based on the distance between the two coordinates by a logistic function so that two coordinates with the greater distance have a higher probability of adding an indexing node to the non-empty point.

(5) Randomly sample a real number according to the probability of adding an indexing node to determine whether intersection links should be added between $c_i$ and $c_j$.

(6) Add an indexing node pij and connect it to ci and cj respectively through partial order relation like inclusion link.

The advantage of using Mahalanobis distance is that the correlation matrix in the distance function can make two vectors orthogonal and normalized with respect to the distribution of the level and the number represented in the weight vector. The probability of adding an indexing node to connect two coordinates at two dimensions control the number of links.   

\subsection{Coordinate Distance}

For a coordinate $c_i$ at dimension $X_i$, a weight vector $w_i =<l_i/L_i, |c_i|>$ is composed, where $l_i$ is the level of $c_i$ at the coordinate tree of $X_i$, $L_i$ is the maximum level of the coordinate tree of dimension $X_i$ and $|c_i|$ is the number of resources under $c_{is}$. Then, a Mahalanobis distance $d(c_i, c_j)$ between two coordinates $c_i$ at dimension $X_i$ and $c_j$ at dimension $X_j$ is calculated based on the weight vectors $w_i$ and $w_j$ of the two coordinates as follows:
\begin{equation*}
D_M\left(c_i,c_j\right)=\ \sqrt{\left(w_i-w_j\right)^TS^{-1}\left(w_i-w_j\right)}\dots\left(1\right)
\end{equation*}

where $\left(w_i-w_j\right)^T$ is the transpose of the difference between two weight vectors $w_i$ and $w_j$,  $S^{-1}$ is the inversion of the correlation matrix of the vector elements of $w_i$. 

The distance $D_M\left(c_i,c_j\right)$ is larger when there is a bigger difference between the levels and the number of resources of two coordinates. Ideally, the correlation matrix should be built from available distributions of $l_i/L_i$ and $|ci|$. However, the two distributions do not directly reflect the intersection size of two coordinates. Moreover, it requires $S$ to be a full rank matrix, which requires the sample set to be large enough and discriminative enough. As each node on the tree structure of a given dimension $X_i$ is unique, the data on $l_i/L_i$ and $|c_i|$ at any given dimension is actually fixed with only one sample, which is unable to construct $S$. Thus, the sample data for $l_i/L_i$ and $|c_i|$ is constructed from all coordinates of $X_i$. That is, when adding a resource $r$ with $r(p) = \{c_1, c_2, ..., c_n\}$, for a coordinate $c_{is}$ on the ancestor tree $Path_i$ of $c_i$, a sample $<s/L_{is}, |c_{is}|>$ is constructed with $L_{is}$ is the depth of the path containing $c_{is}$ on $Path_{i}$ and $|c_{is}|$ is the number of resources under $c_{is}$. Then, each coordinate on $Path_i$ is a sample for obtaining $S$ and $S^-1$. $D_M\left(c_i,c_j\right)$ can be calculated by $S^-1$ for all coordinates within the dimension $X_i$ and $X_j$. 

\subsection{Determine Probability of Adding an Indexing Node}
After determining the distance, a probability of adding an indexing node $p_{} =<c_i, c_j>$ between $c_i$ and $c_j$ (selected when traversing the two dimensions one after another) is calculated as follows: 
\begin{equation*}
P(p_{i,j}) = \frac{1}{1+e^{-D_M\left(c_i,c_j\right)}}\dots\left(2\right)
\end{equation*}

The probability is determined by a logistic function ensure that the greater the distance between the two coordinates, the higher probability of adding a new index. That is, when a new point is added, $P(p_{i,j})$ is calculated and sampled out by a random variable to decide if an intersection indexing node and links are added. The detailed analysis of using Mahalanobis distance for adding intersection links is given in Appendix 3.

Indexing nodes and links are added when a resource at point  $p= <c_1, ..., c_n>$ is inserted. For each coordinate $c_i$, a path $Path_i$ of inclusion links from the root node $T_i$ to $c_i$ is obtained. Then, for two dimensions $X_i$ and $X_{i+1}$, two paths $Path_i$ and $Path_j$ are obtained.  The probability is computed between each coordinate $c_i$ in $Path_i$ and each $c_j$ in $Path_j$. If the sampling random number is larger than the probability, a new indexing node with ID $p_{i,j}=<c_i, c_j>$ is added and two links are added from $c_i$ and $c_j$ to $p_{i,j}$. The resource point $p$ is also linked to $p_{i,j}$. Then, a query visiting indexing node $c_i$ and $c_j$ can follow the two intersection links to reach the index node $p_{i,j}$ and the linked resource point $p$.

The covariance matrix $S$ of the distance function in equation (1) is calculated by the following steps: (1) it collects the weight matrix $D$ that contains vectors of the level and the number of resources of all coordinates of two dimensions $X_i$ and $X_j$; (2) subtract the matrix $D$ from the mean vector $\mu$ of $D$ to obtain the centered data matrix $M$; and, (3) calculate $M_T\times{}M$ to obtain the covariance matrix $S$ and use the inverse operator to obtain $S^-1$.

The calculation cost can be large when all coordinates are used in $D$. To reduce the cost, the longest and shortest path that contain the maximal number of resources and the minimal number of resources are selected as samples for calculating $S$ as they represent the largest variance among weight vectors.

The Mahalanobis distance \textit{D$_{M }$}is a statistically normalized variable
by using the inverse covariance matrix \textit{S}$^{-1}$$_{ }$as weights to
normalize different scales of dimensions. But it is not a normalized distance
within [0, 1]. As the square of the distance from a vector to the mean vector
follows a \textit{d}-dimensional Chi-Square distribution
${\chi{}}^2\left(d\right)$, the significance level $\alpha{}$ can be determined
by the Chi-Square percentile. That is, with \textit{d} degrees of freedom and 
$\alpha{}$ = 0.05, the threshold of the variable is estimated by
${\chi{}}_{0.95}^2$(\textit{d}). For example, for \textit{2}-dimensional data,
${\chi{}}_{0.95}^2\left(2\right)\approx{}5.99$, which means that \textit{v} can
be deemed as an outlier with confidence of 0.95 when $D_M^2\left(v,\mu{}\right)$
is larger than 5.99. This feature can be leveraged to add intersection links when
the coordinate\textit{ c} is deemed as an outlier by the Mahalanobis distance
\textit{D$_{M}$}(\textit{c}, $\mu{}$)$>$$\ {\chi{}}_{0.95}^2$ (\textit{d}).

However, when there are few such outlier coordinates to build intersection
links, the index is too small to work. To control the number of intersection
links, the Mahalanobis distance \textit{D$_{M}$}(\textit{w$_{i}$},
\textit{w$_{j}$}) is used as the variable \textit{x} of a logistic function to
produce a probability $P\left(x\right)=\frac{1}{1+e^{-x}}$within [0, 1], where
\textit{x }= \textit{D$_{M}$}(\textit{w$_{i}$}, \textit{w$_{j}$}) in equation
(2). As \textit{x} becomes large, the value of \textit{P}(\textit{x}) approaches
1, which ensures that the larger the distance, the higher probability of building
an indexing node \textit{p$_{i,j}$} for two coordinates \textit{c$_{i}$} and
\textit{c$_{j }$}in equation (2).

As each link is sampled independently, the expected number
\textit{S}(\textit{n}) of links to be added when the distance is from 0\textit{
}to \textit{n} is:
$E\left(S\left(n\right)\right)=\int_0^n\frac{1}{1+e^{-x}}dx=ln\left(e^n+1\right)-C\approx{}n$when
\textit{n} becomes large.

That is, for those pairs with large distance, a link is almost always added with
high probability. However, this can lead to too much indexing nodes. To make the
number of links below \textit{n}, for example \textit{n}$^{0.5}$, the probability
\textit{P}(\textit{x}) can be changed to:

\[
P\left(x\right)=\frac{e^x}{2\left(e^x+1\right)\sqrt{\ln{\left(e^x+1\right)}}}\dots\left(3\right)
\]

Then, the expected number of links is

\[
\int_0^n\frac{e^x}{2\left(e^x+1\right)\sqrt{ln\left(e^x+1\right)}}dx=\\
\sqrt{\ln{\left(e^n+1\right)}}-\sqrt{\ln{\left(2\right)}}\dots\left(4\right).
\]

and is $\approx n^0.5$ when $n$ is large.

When there are $n$ coordinates in one dimension, the possible intersections can be $n^2$ for two dimensions and the expected number of links is $O(n)$. Obviously, one can use:

\begin{equation*}
  P(x) = \frac{e^x}{a\left(e^x+1\right){(ln{\left(e^x+1\right)})}^{1-\frac{1}{a}}}
\end{equation*}

to produce links with expected number as follows:

\begin{equation*}
E\left(S(n)\right)=\int_{0}^{n}\frac{e^{x}}{a\left(e^{x}+1\right)\left(\ln\left(e^{x}+1\right)\right)^{1 - \frac{1}{a}}}dx=\\
\left(\ln\left(e^{n}+1\right)\right)^{\frac{1}{a}}-C
\end{equation*}

and is $\approx n^{\frac{1}{a}},\quad a > 1$

Fig. 6 shows the algorithm for building the graph index when a new node is added, where lines 1-2 add partial order links for each coordinate ci of the node, lines 3-20 add links to connect coordinates at different dimensions, where line 13 calculates a distance between two coordinates at two dimensions, line 15 calculates the probability of adding an indexing node to connect the two coordinates, lines 17-29 determine whether adding the link by a random sample or not, and line 21 returns the index G.

\begin{figure}[!t]
  \centering
  \includegraphics[width=0.8\columnwidth, keepaspectratio]{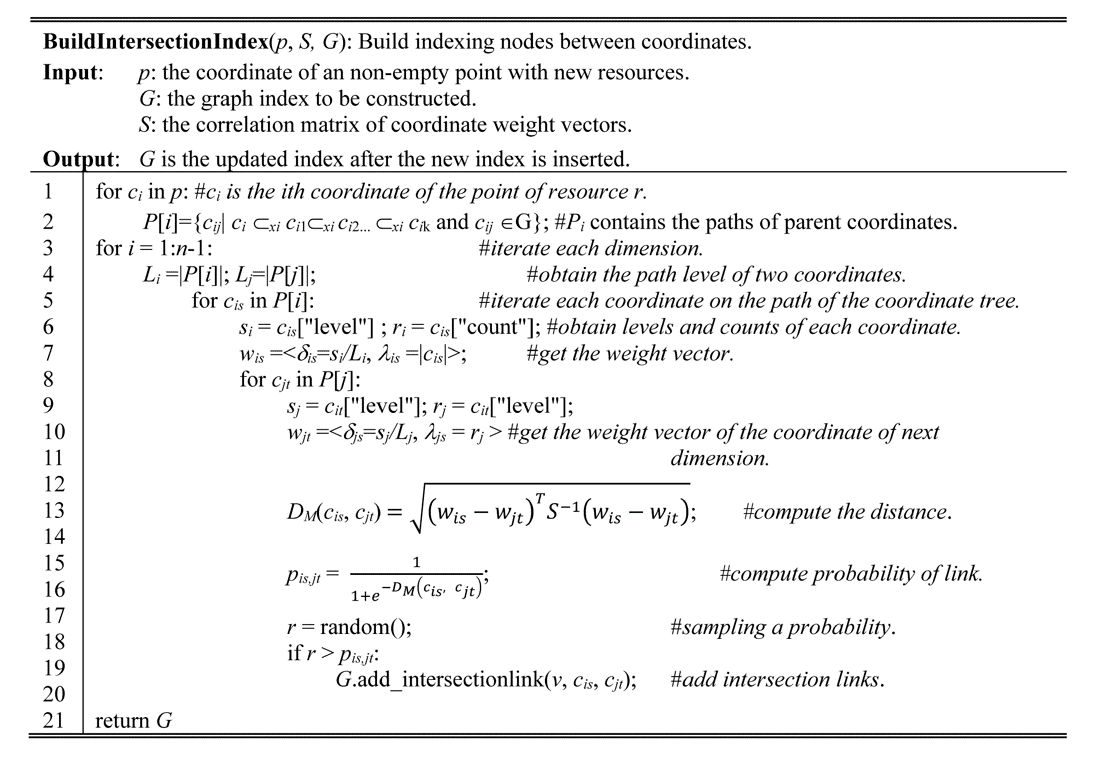}
  \caption{Creating indexing nodes to connect two tree indexing nodes with a probability in proportion to the distance of the two nodes.}
  \label{fig_6}
  \end{figure}

\section{ADDING SHORTCUT LINKS AND SPLITTING NODES}
\subsection{Adding Shortcut Links}
Shortcut links are added to partial order relations between coordinates at a dimension so that a range can be quickly located by jumping from one coordinate to another at the dimension. 

A subspace query following partial order relation can be completed within logarithmic scale when the tree index has a logarithmic scale of depth with respect to the number of points. Fig. 7 shows an example of shortcut links (in red color) added between year coordinates, between season coordinates and between month coordinates at the date dimension.

\begin{figure}[!t]
  \centering
  \includegraphics[width=0.8\columnwidth, keepaspectratio]{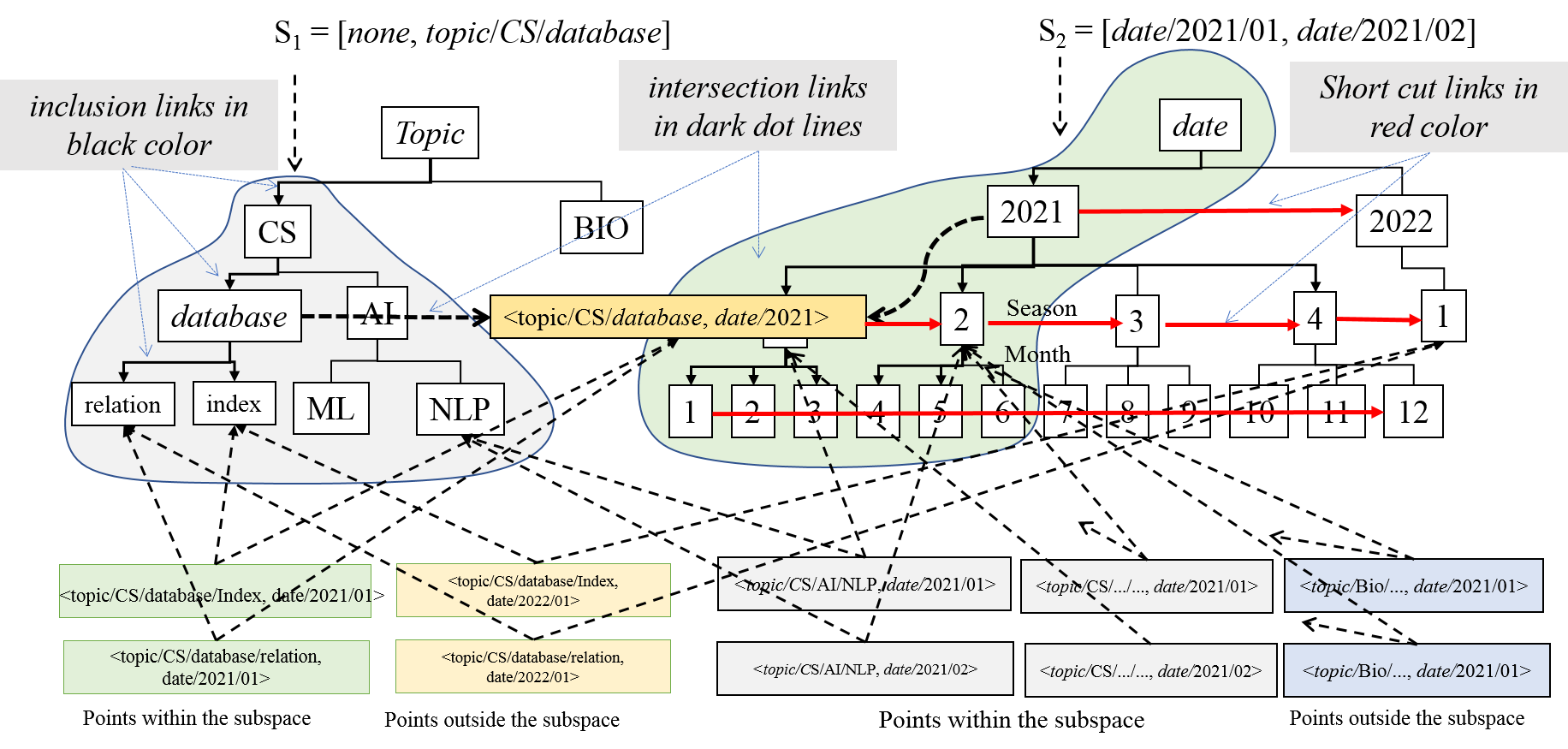}
  \caption{Add a shortcut link on partial order relations of coordinates at date dimension.}
  \label{fig_7}
  \end{figure}
Fig. 8 shows the algorithm for adding shortcut links between coordinates at a dimension, where lines 1-15 scan each dimension, lines 2-6 obtain the tree index $Tree_i$ of a dimension and the coordinate level on Treei, lines 7-15 add shortcut links for each node $v$, lines 8-9 obtain the predecessor and successor of each coordinate $v$ at each level $l$, lines 12-17 add short links for predecessor and successor to $v$, and line 16 returns the updated graph index $G$.
\begin{figure}[!t]
  \centering
  \includegraphics[width=0.8\columnwidth, keepaspectratio]{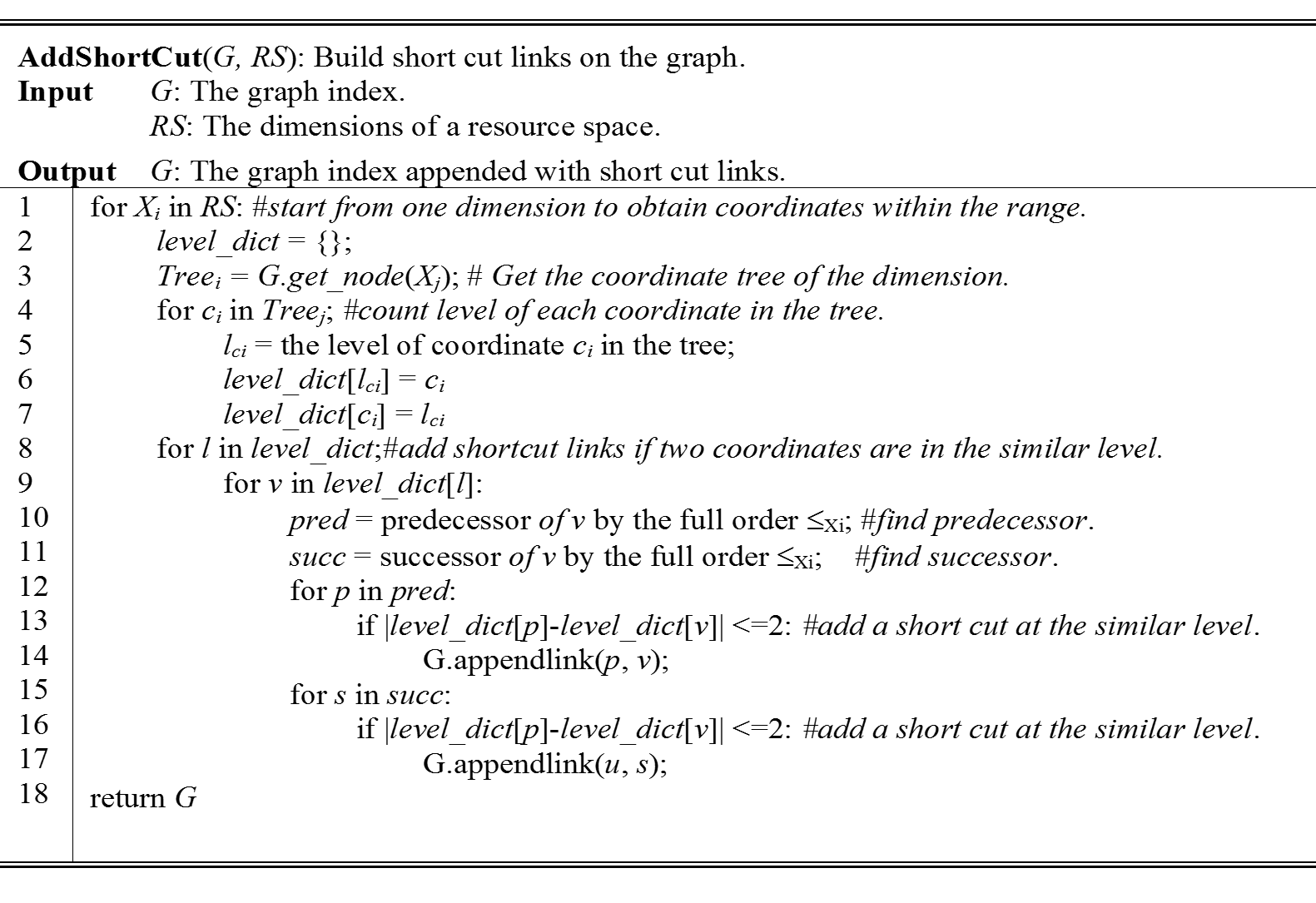}
  \caption{Adding shortcut links.}
  \label{fig_8}
  \end{figure}

\subsection{Splitting Nodes}
With accumulation of resources, the nodes holding big number of resources need to be split to make balance between nodes.  When a new resource at point $p$ under the index node $v$ is added, a probability of $\beta_v =1-1/\left|S_v\right|$ is calculated to determine whether split $v$ or not, where $|S_v|$ is the number of child index nodes of $v$ and $\beta_v$ is the splitting rate to determine how quickly a split operation is conducted. When $|S_v|$ increases, the splitting probability $\beta_v$ decreases. 

When $v$ is to be split, a probability distribution $P_v = {|R(s_i)|/|R(v)|}$, where $s_i$ is the ith child index node of $v$ for $i =1,\dots, k$ is used to make a sample to select a child node $s_i$ of $v$ to conduct split operation, where $|R(v)|=\sum_{s\subset{}v}{|R(s)|}$ is the total resource count under $v$ and $|R(s_i)|$ is the number of resources of $s_i$. 

When a child index node $s_i$ of $v$ is selected based on $P_v$, new index nodes can be built under $s_i$ and resources under index node $s_i$ will be relocated to new index nodes according to their coordinates.  To split $s_i$, a point $p_r$ under $s_i$ is randomly selected and one coordinate $c_k$ of pr at dimension $X_k$ and its sub-coordinates in dimension $X_k$ are used as child index nodes of $s_i$ to split $s_i$.

When a new resource at point $p$ is added, the distribution $P_v(S_v)$ of $v$ is obtained and is used to make a sample to select a child node $s_i \in S_v$ as the node to be split. If $s_i$ is an indexing node with resources, a new indexing node $s$ is added such that $s_i$ is linked to $s$ and $s$ is linked to $v$ (i.e., $s_i \subset s \subset v$). If $s_i$ has no resources but only child indexing nodes, repeat the process by treating $s_i$ as $v$ to check if $s_i$ needs to be further split. To control splitting, a dumb sub index $\empty$ is added to $v$ with the probability of $\beta_v$ being selected by the sampling process and no action is taken when $\empty$ is selected. $\beta_v\ =1-1/{|S_v|}^\alpha$ indicates that when the number of sub index increases, the splitting probability of splitting decreases. $\alpha$ is the splitting rate to determine how quickly an indexing node is conducted.

When an indexing node in $p=<s_1, ..., s_k, s_{k+1},..., s_n>$ is selected to be split by $v = <s_1, ..., s_k>$, a coordinate $c$ from $p \backslash v$ is selected as a new coordinate $s_{k+i}$ together with original $v$ to compose a new indexing node $s =<s_1, ..., s_k, s_{k+i}>$, then link all points under $v$ to $s$ and link $s$ to $v$. 
$s_{k+i}$ is selected according to the resource counts of $s_{k+i}$  at coordinates $s_k$, $s_{k+1}$,..., and $s_n$ of $p$. As $p$ is also linked to $<s_{k+1}>$, ..., and $<s_n>$, the resource count can be obtained and the one with the maximum number of resources is selected as the new coordinate to compose $s$. 

For a given indexing node $v = <c_{d_1}, s_1, c_{d_2}, s_2, \ldots, c_{d_k}, s_k>$ with k dimension coordinates and $c_{d_i}$,$s_j$ being the coordinate $s_j$ of dimension $d_i$. Two types of sub indexing nodes can be linked to $v$ by inclusion links. A resource point $p =<c_{d_1}, s_1, c_{d_2}, s_2, \ldots, c_{d_k}, s_k, c_{d_{k+1}},s_{k+1}, \ldots, c_{d_n},s_n> \subset v$ is connected to $v$ with the number of resources $|R(p)|$. If an indexing node $s \subset v$ is connected to $v$, $R(s) = \sum_{t \subset s}|R(t)|$.  $|R(v)| = \sum_{t\subset s}|R(t)|$ is the total number of resources under $v$ and each coordinate $s \subset v$ occupies $|R(s)|/|R(v)|$ of resources under $v$. Let $S_v ={s_i|s_i \subset v for i=1, \ldots, k}$, then a distribution $P_v(S_v)$ of resources in $v$ can be obtained as $P_v(S_v) = {|R(s_i)|/|R(v)| for i =1, \ldots, k}$. If $S_v$ evenly partitions $R(v)$, the indexing node $v$ is still balanced. If there is a large variance among coordinates of $v$, an adjustment can be made by partitioning the coordinate with large number of resources by a set of coordinates following a coordinate of another dimension.

\begin{figure}[!t]
  \centering
  \includegraphics[width=0.8\columnwidth, keepaspectratio]{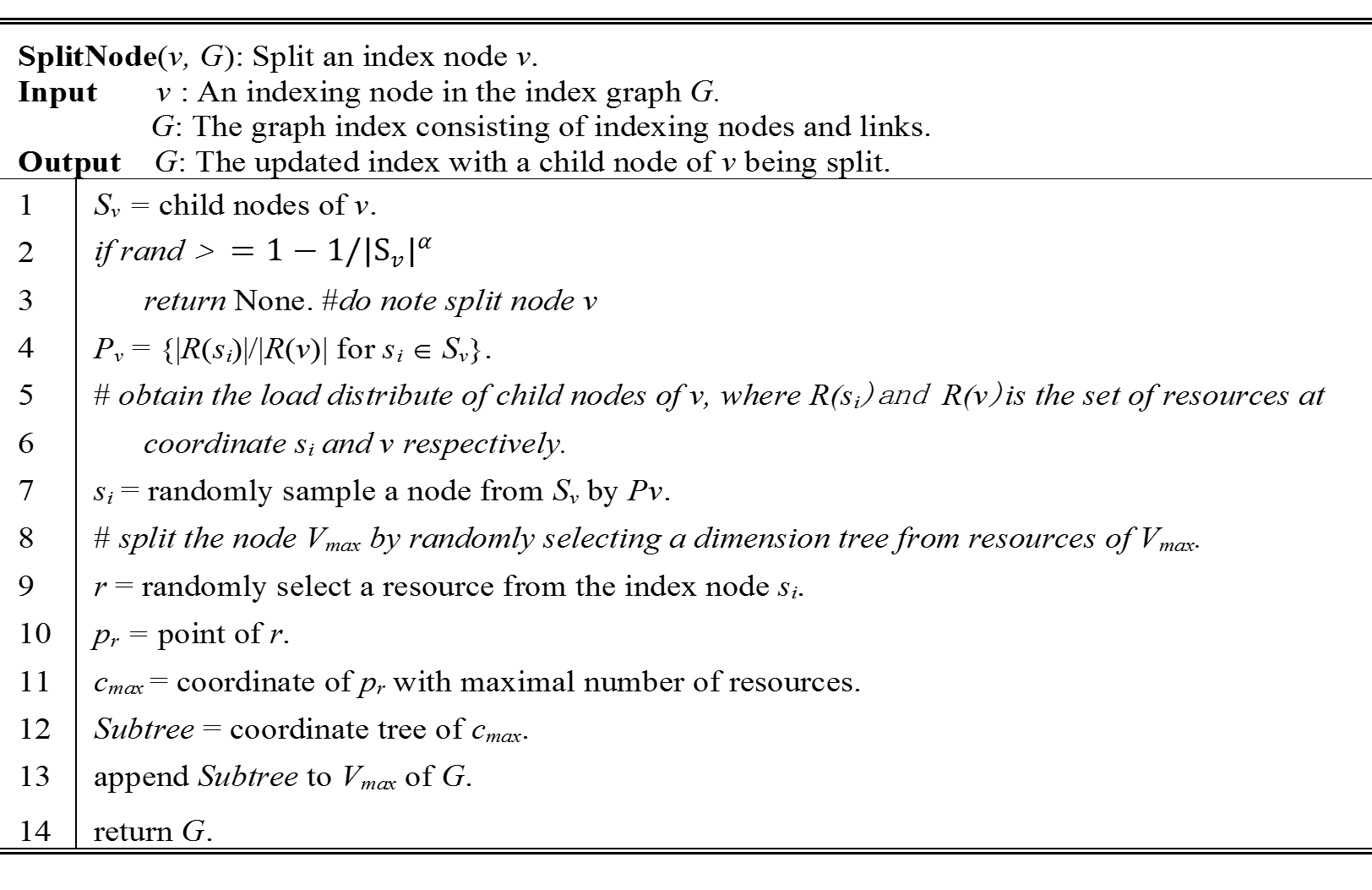}
  \caption{Splitting an indexing node.}
  \label{fig_9}
  \end{figure}

Fig. 9 shows the algorithm for splitting an indexing node if one of its child indexing nodes holding more resources than others, where  line 1 obtains all child indexing nodes of the indexing node $v$, line 2 obtains the distribution $P_v$ of number of resources at child indexing nodes of $v$, lines 5-8 obtain the child indexing nodes with the maximal number of resources and the minimal number of resources according to the distribution of number of resources, lines 9-15 split the indexing node $V_max$ with maximal number of resources if lmax is two times of the minimal load lmin in $P_v$, line 11 randomly selects a resource at $V_max$, line 12 obtains the coordinates of the resource and selects one coordinate $c_max$ and adds its subtree to the indexing node $V_max$, and line 16 returns the updated index graph $G$.

\section{GENERATING INDEX}

As key-value database can efficiently store and query with unique keys \cite{RN9}, using coordinates of points as keys to store resources in a key-value database is a way to efficiently  store resources and query with coordinates of points. A subspace query can be mapped into multiple key-value queries with the coordinates of points in the subspace as keys. 

\begin{figure}[!t]
  \centering
  \includegraphics[width=0.8\columnwidth, keepaspectratio]{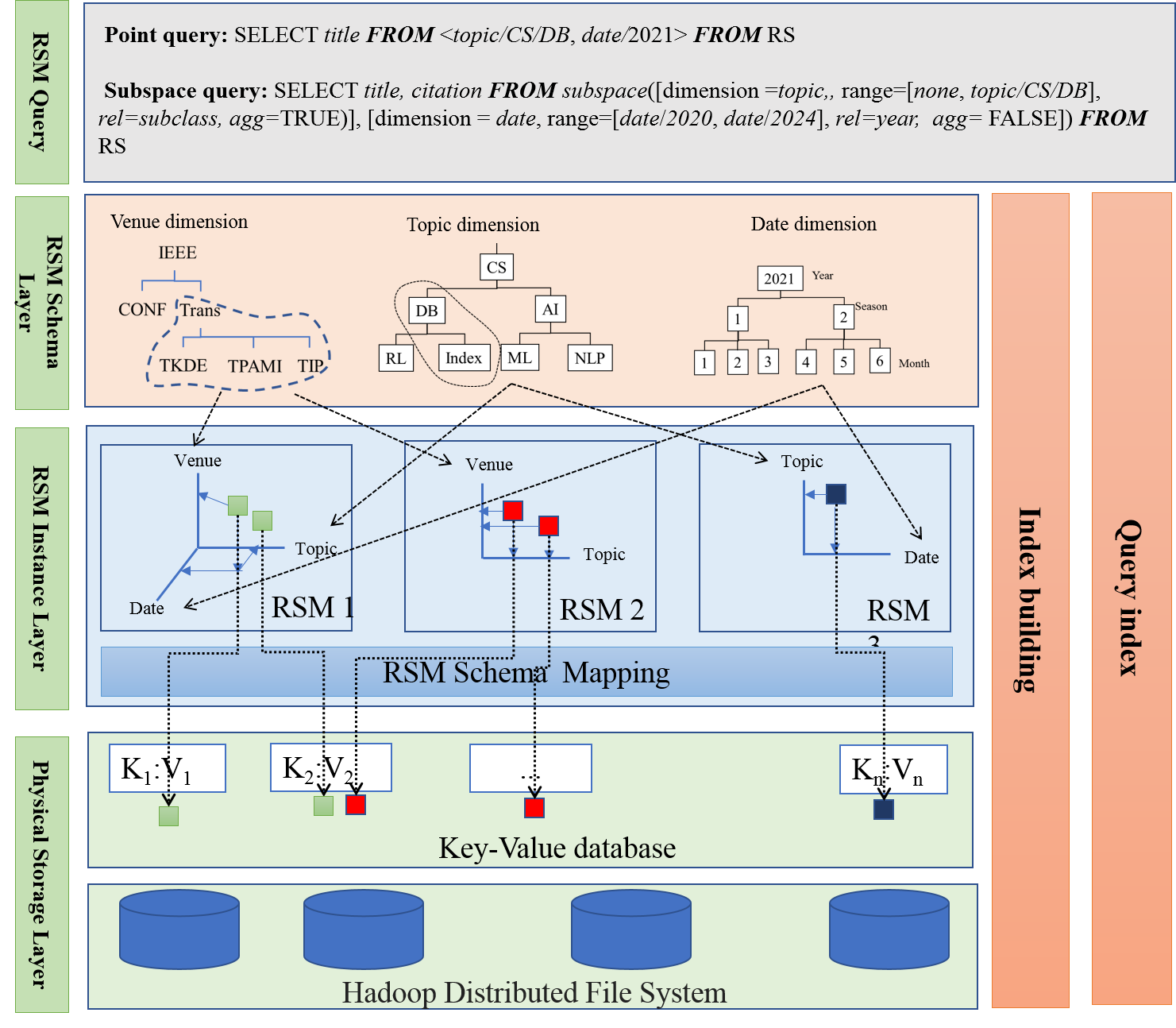}
  \caption{Architecture for storing resource space.}
  \label{fig_10}
  \end{figure}

Fig. 10 is the system framework for storing a resource space (RS) in a key-value database. The underlying layer is the key-value database deployed on a distributed file system. Multiple instances of resource space can be stored on the key-value database as the instance layer. The schema layer defines the structure of resource space. The top layer is the query interface. The right part is the index structure to support efficient subspace query with aggregation by mapping a query into points in subspace and keys of the underlying key-value database to obtain resources.

Locating a resource in $p$ by its ID requires to find the resource set $R(p)$ first by calling a key-value query on the key-value database using p as the key with $O(1)$ time and then finding an ID from $R(p)$ with another $O(|Rp(p)|)$ steps. 

However, key-value database does not consider the partial order relations of coordinates for processing a subspace query with aggregation. For a point $p =<c_1, ..., c_n>$ and a subspace query on RS(S) = $S_1\times ...\times S_n$ with $S_i = [l_i, u_i]$, it needs to check if $l_i \subset_{X_i}  c_i \subset_{X_i}  u_i$ for determining if $p$ is in the query result. A reachability matrix can be built for $\subset_{X_i}$ so that $l_i \subset_{X_i}  c_i \subset_{X_i} u_i$ can be immediately checked, it still needs $N \times D$ comparisons for $N$ resources with $D$ dimensions to obtain resources within the range, where many points outside the range needs to be checked. To facilitate a subspace query on multiple dimensions, a resource at $p = <c_1, c_2, ..., c_n>$ can be mapped into multiple key-value pairs by using each coordinate $c_1$, $c_2$, ..., and $c_n$ as a key. 
A subspace query can be decomposed into an intersection calculation of a set of sub-queries on each dimension, but storing resources at each of its coordinates can make resource set at each coordinate too large to calculate intersection. An index is generated to index resources at points of a subspace to their keys in the key-value document database to help efficiently locate points within a subspace with a lower cost of calculating intersection according to the inclusion and partial order relation on points.  
In summary, a graph index is generated by giving a resource space RS with a set of resources by the following steps:
(1) Define dimensions of RS in an XML file.
(2) Match each resource with a point in the RS, store the point as key and the resource as value into a key-value database, and then unite the resources sharing the same point as the value of the point, represented as $(Key=point, Value=R(p))$.
(3) Call the algorithm for generating index presented in Fig. 2 with RS in form of XML and the key-value database to create a graph index G (including the procedures for the index tree generation, indexing nodes generation between coordinates, nodes splitting and shortcut links adding).
(4) Store G in a JSON file for query usage.
(5) Process query and new resources: When a subspace query is received, the algorithm presented in Fig. 4 for processing the query is called to generate the answer. When a set of new resources is added, perform step (2) to add each new resource into the database. 

\section{SUBSPACE QUERY LANGUAGE}
A SQL-like subspace query statement is adopted to facilitate subspace query with aggregation. It consists of the specifications of selection operator, top-k operator, function call operator and subspace query operator, and each operator is represented by a FROM sub-statement, represented as follows.

\small
\begin{verbatim}

SELECT [point_dimension|point_variable]+, 
     [resources_attribute]  
FROM TOP_RESOURCE(TOPK=n_1,  
     MEASURE=[resource_attribute|esource_variable])
FROM,TOP_POINT(TOPK=n,
     MEAURE=[point_dimension|point_variable])
FROM SUBSPACE([
	dimension = dimension_name,
	range = [a, b],
	rel = relation_name,
	agg = TRUE|FALSE,
	[resource_variable = func(resource_attribute+)]*,
	[point_variable = iterate_func([resource_coordinate|resource_attribute]+)]*
	]+)
FROM RS;
\end{verbatim}
\normalsize

The $SELECT$ operator specifies the coordinates and attributes of resources to be displayed as the output of the result. The $TOP_RESOURCE$ is an operator for selecting resources with top-k values of an attribute or a variable from each point of the query result. The $TOP_POINT$ is an operator for selecting points with top-k values of coordinate or a calculated variable from the query result.  The $SUBSPACE$ operator defines aggregation ranges at dimensions.

\section{EXPERIMENTAL EVALUATION}

Fig. 11 shows two different probability functions, where the red one is the function in equation (2) and the blue line is the function in equation (3). As variable x increases, the probability of adding links decreases according to equation (3), which makes the total number of links to be added under n.  A set of 100 coordinates is randomly produced from levels 1 to level 14 and the number of resources from 0 to 1000 for two dimensions. Then, the two dimensions will have 104 possible intersections. According to (1) and (2), the probability of adding a link is shown in the blue curve in Fig. 11, which produces 8300 expected links and the green curve corresponds to equation (3) with the expected number of links 1554. The observed expected number of links is much smaller than 104 because the Mahalanobis distance follows a normal distribution with a small portion of large distance values, which makes the expected number links even smaller.
An experiment is conducted to verify the subspace query performance on the index. The ACM CCS category tree used as a topic dimension has wide width but shallow depth. The distribution of depths is shown in Fig.12. A two-dimensional RS instance with date dimension and topic dimension is built and a set of randomly produced resources are added to the RS instance to evaluate the query performance. A TF-IDF based method (denoted as tfidf) and the index with only coordinate trees (denoted as tree-only) are compared with the graph index (denoted as graph). The tfidf method queries point by keywords of coordinates. The tree-only method uses keywords of paths on tree index to locate points. Fig. 13a shows that the graph index can reduce the match counts (the number of comparisons between a query and the coordinates of a point) in querying processing. Indexes with shortcut links and splitting nodes are also compared in Fig. 13b.
To demonstrate the efficiency of the graph index structure, a set of subspace queries is conducted on the index with indexing nodes on non-empty points and shortcut links within tree index to show how those links can help improve the efficiency of query. The results are shown in Fig. 13. In many cases, creating indexing nodes on non-empty points and build shortcut links for each tree indexes can help reduce the times of comparisons.

  \begin{figure}[!t]
    \centering
    \includegraphics[width=0.8\columnwidth, keepaspectratio]{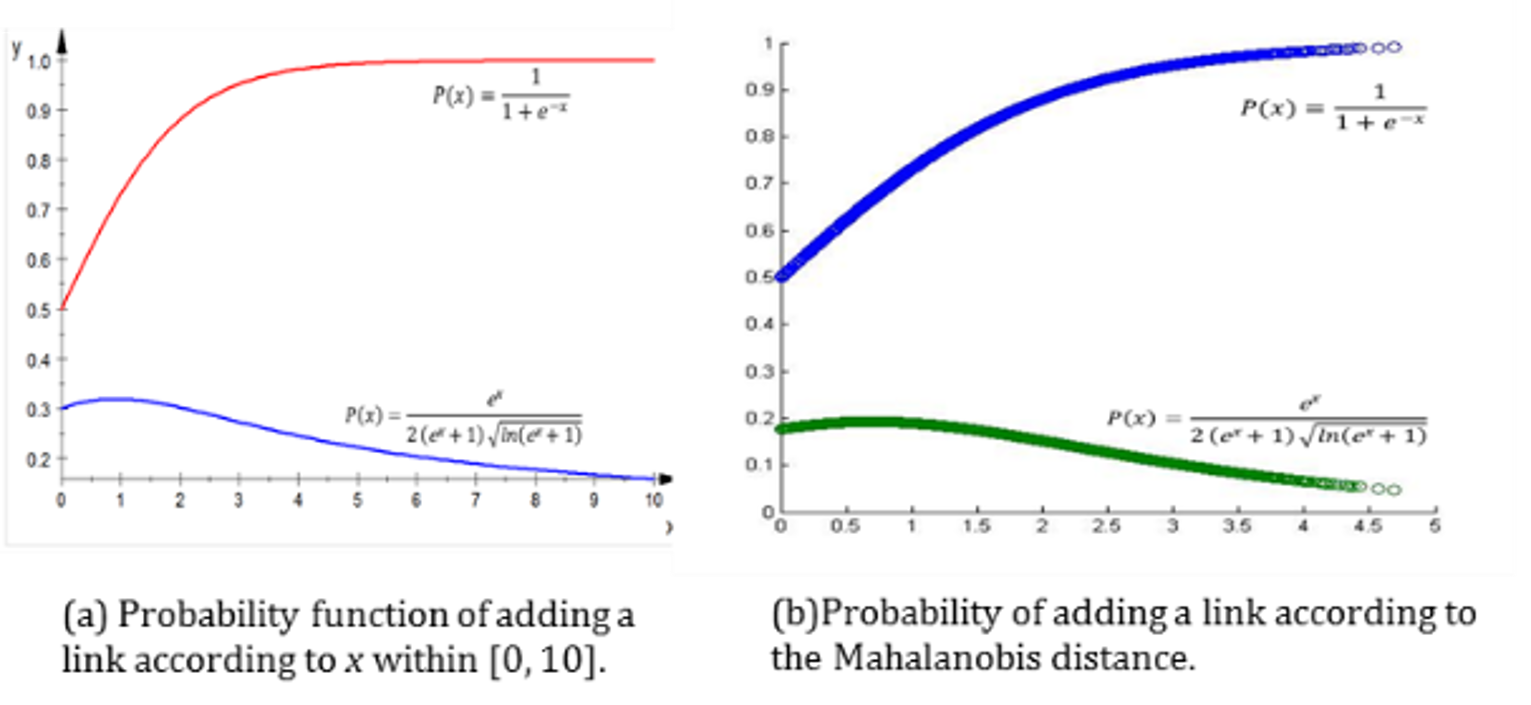}
    \caption{ Probability function of adding a link according to x within [0, 10] and the Mahalanobis distance.}
    \label{fig_11}
    \end{figure}
\begin{figure}[!t]
  \centering
  \includegraphics[width=0.8\columnwidth, keepaspectratio]{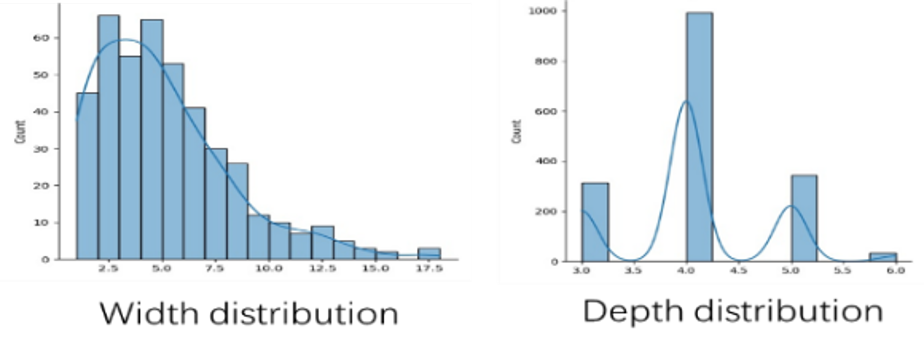}
  \caption{ Coordinate tree distribution.}
  \label{fig_12}
\end{figure}
\begin{figure}[!t]
    \centering
    \includegraphics[width=0.8\columnwidth, keepaspectratio]{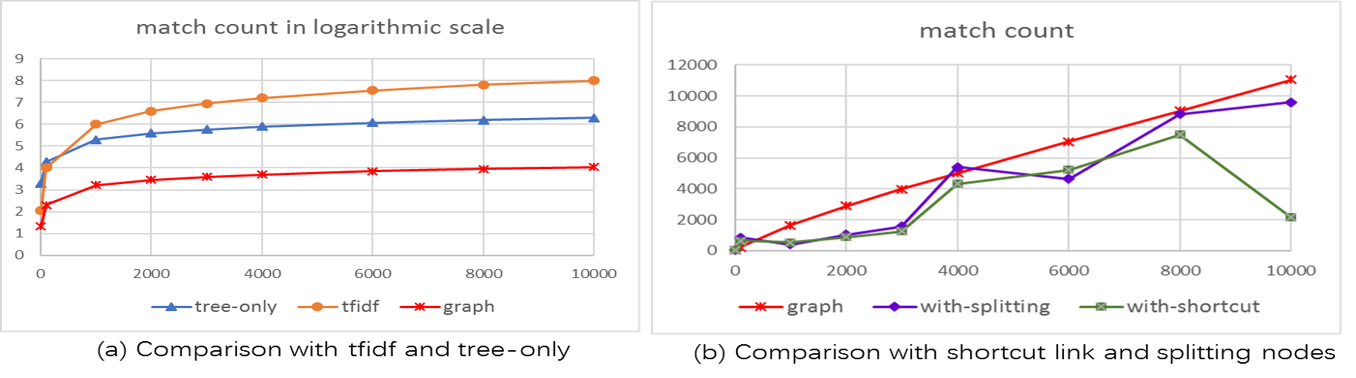}
    \caption{Match count of different approaches for querying random points.}
    \label{fig_13}
\end{figure}
\begin{figure}[!t]
      \centering
      \includegraphics[width=0.8\columnwidth, keepaspectratio]{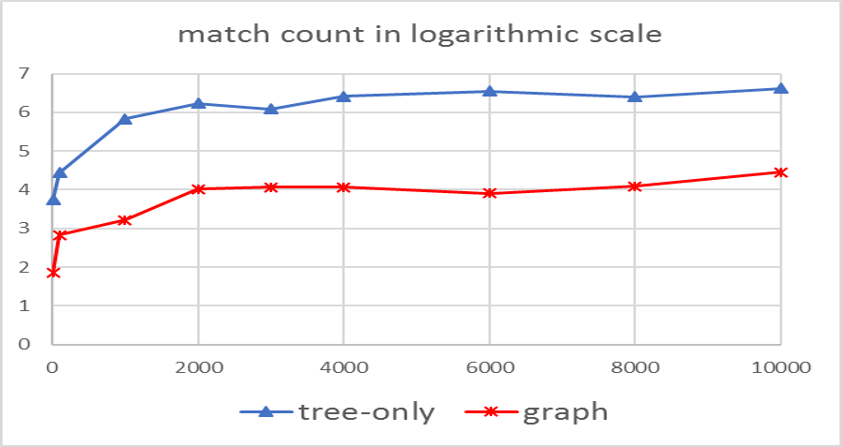}
      \caption{Match counts increasing ratio in a 3-dimensional resource space.}
      \label{fig_14}
\end{figure}
        
A 3-dimensional resource space is used to test the cost of query count when the number of dimension increases. As indexing nodes are created to connect two dimensions, a query follows the links to quickly narrow the subspace while the keyword-based method still needs to compare keywords at all dimensions. Fig.14 shows the logarithmic match counts of queries when using inclusion links between indexing nodes in a three-dimensional resource space. The index can achieve a better query performance than using keywords to query.

\section{RELATED WORK}
\subsection{Relational Data Model}
Classical relational database efficiently implements queries on attributes based on functional dependencies \cite{RN10}. 
Traditional multidimensional data models regard a given set of attributes of data items as dimensions and values of the attributes of data items as coordinates of dimensions. Data Warehouse Online Analytics (OLAP) implements hierarchical aggregation query on multi-dimensional continuous numerical data \cite{RN11}. Different from relational databases, the coordinates of a dimension are organized in a hierarchical structure such that the value at a coordinate can include values from its sub-coordinates. This hierarchical structure cannot be represented by classical relational data model. 
Data cube is to process multidimensional query on data through quantitative measurement. The classic data cube model in OLAP provides roll up and drill down query operators to aggregate data values from coordinates of lower levels to coordinates of higher levels. Text-cube generalized data cube by allowing text sets to be stored in a multidimensional data cube where partial order relation is defined between attributes \cite{RN12}. These works do not support aggregation operation at dimensions with partial order on coordinates. 
In general, most multidimensional data models focus on managing and exploring values in a multidimensional space. Only quantitative measurements can be kept and aggregated in multidimensional data query.  It is still a challenge to implement range query and aggregation operation in a multi-dimensional data space. 

\subsection{Multidimensional Index}
Building multidimensional indexes is an effective way to support range query and neighborhood query in low-dimensional data space \cite{RN13}.  Classical R-tree extends the idea of B+-tree to the d-dimensional space, using minimum bounding box to store multidimensional data points, and building a tree index using the containment relation between minimum bounding boxes to improve the efficiency of multidimensional data interval queries. An R-tree indexing approach based on the non-overlapping intervals can enhance the efficiency of data insertion and querying \cite{RN14}. K-D tree uses data points to build hyperplanes and divides the space by hyperplanes to form a balanced binary tree index \cite{RN15}. Due to the uneven distribution of point positions, hyperplanes can be uneven, which leads to the decrease of the efficiency of range queries.  To solve the problem of uneven K-D tree, QuadTree divides each dimension into two intervals then two dimensions form a hyperplane with four equal-sized sub-hyperplanes \cite{RN16}. Each hyperplane containing more than one node can be further divided so that the subspace with more data can be split into more subspaces to form a more balanced tree index. However, those classical indexes cannot handle high dimensional data. Using small-world graph to index high dimensional data for approximate nearest neighborhood query can efficiently handle uneven data distribution issues in high dimensional data \cite{RN17}. Bitmap based index helps quickly locate ranges at dimensions by mapping ranges of a high dimensional space into a 2-dimensional index \cite{RN18}.  
Classical range query operator is defined in a multidimensional data space to obtain data items within the range supported by various index built on a space distance metric for partitioning data sets into hierarchical groups on indexing nodes \cite{RN19}.  Range query in distributed system can aggregate resources from different peers in the system with a certain topology and data location schema \cite{RN20}, but it lacks dimension specification with hierarchical coordinates at dimensions.  LMS index has been widely studied and used in key-value database to support key-range query by storing key-range in a logged memory and merged in a tree structure when necessary \cite{RN20}. How to implement resource aggregation along hierarchical coordinates has not been addressed.  
Multi-dimensional space indexes built on distance metrics can avoid traversing all data to achieve high efficiency for neighborhood query in linear high-dimensional spaces \cite{RN13}, but it cannot apply to a non-linear multi-dimensional space consisting of discrete data and discrete attributes. Most indexes on quantitative data in a multidimensional space with linear coordinates at dimensions cannot handle aggregation operation of discrete resources within a range on a partial order relation between coordinates.

\section{CONCLUSION}
A subspace query with aggregation in multidimensional resource space is proposed to support locating, ranking and aggregating resources in a subspace from a multidimensional resource space with partial order relations on coordinates at its dimensions. 

A graph index generation approach is proposed to efficiently process subspace query with aggregation.  It consists of a tree of coordinates as indexing nodes at each dimension, indexing nodes to non-empty points, and shortcut links within tree of index nodes generated based on partial order relations between points for locating points in a querying subspace with less intersection calculation. Constructing a probability distribution based on the Mahalanobis distance between coordinates at two dimensions and adding index nodes to non-empty points can effectively improve query efficiency while controlling the increasing number of index nodes.  

The randomized node splitting operation based on the depths of child indexing nodes of an index node can make a balanced index. The shortcut links built on partial relations like inclusion can reduce the costs of building index and improve the efficiency of locating ranges within a dimension. The generated graph index is efficient in supporting subspace query with aggregation as it can aggregate resources along the paths on graph index.

The subspace query with aggregation on resources in resource space is important for multidimensional data management as partial order relations such as subclass, inclusion and comparison relations widely exist between various documents and the query can help locate the resources aggregated at different levels of points through the partial order relations to support ranking, selection and calculation on the resources in applications. The graph index can be generated from different partial order relations on dimensions to make the query more expressive and efficient.
Implementing aggregation querying large-scale resources challenges both query expressiveness and efficiency of indexing mechanism.  This work demonstrates the feasibility of implementing efficient query on resource space.

Ongoing work is to extend the partial order relations on coordinates to more relations and apply the approach to providing multi-dimensional abstraction and partial order relation for efficiently managing various resources for domain applications.


\bibliographystyle{IEEEtran}
\bibliography{sn-bibliography}

\end{document}